\documentclass[hyper,a4paper]{JHEP3} 
\usepackage{epsfig}
\usepackage{latexsym}

\usepackage{graphicx}

\usepackage{mathrsfs}
\usepackage{amsmath}
\usepackage{amsfonts}   
\usepackage{amssymb}    
\usepackage{subfigure}

\def\e3{$\epsilon_3$}

\def\co#1{{\ifmmode{\cal O}_{#1}\else${\cal O}_{#1}$\fi}}

\newdimen\unit
\def\point#1 #2 #3{\vbox to0pt{\kern-#2\unit
 \hbox{\kern#1\unit#3}\vss}
\nointerlineskip}

\newcommand{\be}{\begin{equation}}
\newcommand{\ee}{\end{equation}}
\newcommand{\bea}{\begin{eqnarray}}
\newcommand{\eea}{\end{eqnarray}}

\newcommand\ps{\mbox{ ps}} 
\newcommand{\mev}{\mbox{ MeV}}
\newcommand{\gev}{\mbox{ GeV}}

\newcommand{\cl}{\text{CL}}

\newcommand{\alphaemmz}{\alpha_{\text{em}}(M_Z)^{\overline{MS}}}
\newcommand{\alphas}{\alpha_s(M_Z)^{\overline{MS}}}

\newcount\hour
\newcount\minute
\newtoks\amorpm
\hour=\time\divide\hour by60 \minute=\time{\multiply\hour by60
\global\advance\minute by- \hour}
\edef\standardtime{{\ifnum\hour<12 \global\amorpm={am}%
   \else\global\amorpm={pm}\advance\hour by-12 \fi
   \ifnum\hour=0 \hour=12 \fi
   \number\hour:\ifnum\minute<100\fi\number\minute\the\amorpm}}
\edef\militarytime{\number\hour:\ifnum\minute<100\fi\number\minute}
\def\bold#1{\setbox0=\hbox{$#1$}%
    \kern-.025em\copy0\kern-\wd0
    \kern.05em\copy0\kern-\wd0
    \kern-.025em\raise.0433em\box0 }

\newcommand{\newc}{\newcommand}
\newc\eg{{\rm {e.g.}}}  \newc\etal{{\rm {et al.}}} \newc\ie{{\rm i.e.}}
\newc\etc{{\rm {etc}}}
\newcommand\lsim{\mathrel{\rlap{\lower4pt\hbox{\hskip1pt$\sim$}}
   \raise1pt\hbox{$<$}}}
\newcommand\gsim{\mathrel{\rlap{\lower4pt\hbox{\hskip1pt$\sim$}}
   \raise1pt\hbox{$>$}}}
\newc{\mhalf}{m_{1/2}}      \newc{\mzero}{m_0}
\newc{\tanb}{\tan\beta}
\newc{\azero}{A_0}
\newc{\at}{A_t} \newc{\ab}{A_b} \newc{\atau}{A_\tau}
\newc{\bmu}{B\mu}           \newc{\sgn}{{\rm sgn}}
\newc{\mone}{M_1}           \newc{\mtwo}{M_2}

\newc{\charone}{\chi_1^\pm} \newc{\mcharone}{m_{\chi_1^\pm}}

\newc{\hl}{h}               \newc{\mhl}{m_{\hl}}   \newc{\gammahl}{\Gamma_{\hl}}
\newc{\hh}{H}               \newc{\mhh}{m_{\hh}}   \newc{\gammahh}{\Gamma_{\hh}}
\newc{\ha}{A}               \newc{\mha}{m_{\ha}}   \newc{\gammaha}{\Gamma_{\ha}}
\newc{\hpm}{H^{\pm}}        \newc{\mhpm}{m_{\hpm}} \newc{\gammahpm}{\Gamma_{\hpm}}
\newc{\hp}{H^{+}} \newc{\mhp}{m_{\hp}} \newc{\hm}{H^{-}}
\newc{\mhm}{m_{\hm}}
\newc{\xt}{X_{t}}           \newc{\xb}{X_{b}}

\newc{\qzero}{Q_0}          \newc{\qstop}{Q_{\widetilde t}}
\newc{\amu}{a_{\mu}}        \newc{\amususy}{a_{\mu}^{\text{SUSY}}}
\newc{\amuexpt}{a_{\mu}^{\text{expt}}}        \newc{\amusm}{a_{\mu}^{\text{SM}}}
\newc{\deltaamususy}{\delta a_{\mu}^{\text{SUSY}}}
\newcommand{\bsg}{\bsgamma}
\newc\gmtwo{(g-2)_{\mu}} \newc\deltaamu{\Delta a_{\mu}}
\newc\deltagmtwo{\delta (g-2)_{\mu}} 
\newc{\msbar}{\overline{MS}} \newc{\drbar}{\overline{DR}}
\newc{\yt}{h_t} \newc{\yb}{h_b} \newc{\ytau}{h_{\tau}}

\newc{\mtop}{m_t}               \newc{\mtpole}{M_t}
\newc{\mtaupole}{m_{\tau}^{\text{pole}}}
\newc{\mtmtsmmsbar}{m_t(m_t)^{\msbar}_{{\text{SM}}}}
\newc{\mtmtsmdrbar}{m_t(m_t)^{\drbar}_{{\text{SM}}}}
\newc{\mtmtmssmdrbar}{m_t(m_t)^{\drbar}_{{\text{SUSY}}}}

\newc{\mbmbmsbar}{m_b(m_b)^{\msbar} }

\newc{\mbmbsmmsbar}{m_b(m_b)^{\msbar}_{{\text{SM}}}}
\newc{\mbmzsmmsbar}{m_b(\mz)^{\msbar}_{{\text{SM}}}}
\newc{\mbmzsmdrbar}{m_b(\mz)^{\drbar}_{{\text{SM}}}}
\newc{\mbmzmssmdrbar}{m_b(\mz)^{\drbar}_{{\text{SUSY}}}}

\newc{\mtaumzsmmsbar}{m_{\tau}(\mz)^{\msbar}_{{\text{SM}}}}
\newc{\mtaumzsmdrbar}{m_{\tau}(\mz)^{\drbar}_{{\text{SM}}}}
\newc{\mtaumzmssmdrbar}{m_{\tau}(\mz)^{\drbar}_{{\text{SUSY}}}}

\newc{\mgut}{M_{\rm GUT}}
\newc{\mplanck}{M_{\rm P}}      \newc{\mpl}{M_{\text{Pl}}}
\newc{\msusy}{M_{\rm SUSY}}      \newc{\ms}{M_{\text{S}}}
\newc{\jxf}{J({\xf})}
\newc{\jxfexact}{J_{\rm exact}({\xf})}  \newc{\jxfexp}{J_{\rm exp}({\xf})}
\newc{\VEV}[1]{\langle #1 \rangle}
\newc{\xf}{x_f}
\newc\vrel{v_{\rm rel}}
              
\newc\sell{{\widetilde e}_L}      \newc\msell{m_{\sell}}
\newc\selr{{\widetilde e}_R}      \newc\mselr{m_{\selr}}
\newc\snue{{\widetilde \nu}_e}      \newc\msnue{m_{\snue}}
\newc\snutau{{\widetilde \nu}_\tau}      \newc\msnutau{m_{\snutau}}
\newc\supl{{\widetilde u}_L}      \newc\msupl{m_{\supl}}
\newc\supr{{\widetilde u}_R}      \newc\msupr{m_{\supr}}
\newc\sdl{{\widetilde d}_L}      \newc\msdl{m_{\sdl}}
\newc\sdr{{\widetilde d}_R}      \newc\msdr{m_{\sdr}}

\newc\sfermion{\tilde f}  \newc\msfermion{m_{\sfermion}}
\newc\cmeter{{\rm cm}} \newc\meter{{\rm m}} \newc\kmeter{{\rm km}}
\newc\second{{\rm sec}}
\newc{\gstar}{g_\ast}           \newc{\gsstar}{g_{s\ast}}
\newc{\geff}{g_{\rm eff}}
\newcommand\mz{m_{Z}}

\newc{\sthw}{\sin\theta_W}              \newc{\cthw}{\cos\theta_W}
\newc{\bino}{\widetilde B}              \newc{\wino}{\widetilde W_30}
\newc{\higgsinob}{{\widetilde H}^0_b}   \newc{\higgsinot}{{\widetilde H}^0_t}
\newc{\abund}{\Omega h^2}
\newc{\abundchi}{\Omega_\chi h^2}
\newc{\abundcdm}{\Omega_{\text{CDM}} h^2}
\newc{\omegam}{\Omega_{M}}       \newc{\abundm}{\Omega_{M} h^2}
\newc{\omegab}{\Omega_{b}}       \newc{\abundb}{\Omega_{b} h^2}
\newc{\omegacdm}{\Omega_{CDM}}
\newc{\omegatot}{\Omega_{TOT}}
\newc{\rhocrit}{\rho_{crit}}
\newc{\rhochi}{\rho_{\chi}}

\newc\BR{BR}
\newc\bsgamma{b\rightarrow s \gamma }
\newc\brbsgamma{\BR(\overline{B}\rightarrow X_s\gamma)}

\newcommand\brbsmumu{\BR(\overline{B}_s\to\mu^+\mu^-)}

\newcommand\delmbs{\Delta M_{B_s}}
\newcommand\brbtaunu{\BR(\overline{B}_u\to \tau \nu)}

\newc{\beq}{\begin{equation}}
\newc{\eeq}{\end{equation}}

\newc\stoponetwo{{\widetilde t}_{1,2}}
\newc\sbotonetwo{{\widetilde b}_{1,2}}
\newc\stauonetwo{{\widetilde \tau}_{1,2}}

\newc{\sigsip}{\sigma^{SI}_{p}} \newc{\sigsin}{\sigma^{SI}_{n}}
\newc{\sigsiN}{\sigma^{SI}_{N}}
\newc{\sigsdp}{\sigma^{SD}_{p}} \newc{\sigsdn}{\sigma^{SD}_{n}}
\newc{\sigsiA}{\sigma^{SI}_{A}}

\newc\xilim{\xi_{\rm lim}} 
\newc\tlim{t_{\rm lim}} 
\newc\zetalim{\zeta_{\rm lim}} 

\newc\zetah{\zeta_h}
\newc{\relprobone}[1]{p({#1} \vert d)}
\newc{\relprobtwo}[2]{p({#1},{#2} \vert d)}

\long\def\begincomment#1\endcomment{%
       \begingroup\sf\baselineskip12pt#1\endgroup}

\newcommand{\squishlist}{
  \begin{list}{$\bullet$}
   { \setlength{\itemsep}{0pt}      \setlength{\parsep}{3pt}
     \setlength{\topsep}{3pt}       \setlength{\partopsep}{0pt}
     \setlength{\leftmargin}{1.em} \setlength{\labelwidth}{1em}
     \setlength{\labelsep}{0.5em} } }
\newcommand{\squishend}{
   \end{list}  }
        
\newcommand{\nuis}{\psi}
\newcommand{\params}{\theta}
\newcommand{\basis}{m}

\newcommand{\sineff}{\sin^2 \theta_{\rm{eff}}}

\newcommand\mnras[3]   { 
		{{\it Mon.\ Not.\ R.\ Astron.\ Soc.\ }{\bf #1} (#2) #3}}

\newcommand\rpp[3]  {
		{{\it Rept.\ Prog.\ Phys.\ }{\bf #1} (#2) #3}}

\newcommand{\dr}{{\rm d}}

\newc{\chisq}{\chi^2}  \newc{\chisqred}{\chi^2_{\rm red}}

\newc{\ww}{0.49\linewidth} 
\newc{\ttr}{0.31\linewidth} 
\newc{\qq}{0.24\linewidth}

\newcommand{\mdl}{\mathcal{M}}
\newcommand{\predlike}{\mathscr{L}}
\newcommand{\Rr}{\mathscr{R}}
\newcommand{\ltest}{\predlike\text{--test}}
\newcommand{\rtest}{\Rr\text{--test}}
\newcommand{\Hy}{\mathcal{H}}

\newcommand{\data}{d}
\newcommand{\test}{\mathscr{D}}
\newcommand{\dass}{D}
\newcommand{\epem}{e^+e^-}

\title{Are 
{\boldmath $\brbsgamma$ and $\gmtwo$}
  consistent within the Constrained MSSM?} 
	          
\author{Farhan Feroz\\
       Astrophysics Group, Cavendish Laboratory, University of Cambridge \\
       J.J. Thomson Avenue, 	Cambridge, CB3 0HE, UK \\
        E-mail: \email{ff235@mrao.cam.ac.uk}}

\author{Mike Hobson\\
       Astrophysics Group, Cavendish Laboratory, University of Cambridge \\
       J.J. Thomson Avenue, 	Cambridge, CB3 0HE, UK \\
        E-mail: \email{mph@mrao.cam.ac.uk}}

\author{Leszek Roszkowski\\
       Department of Physics and Astronomy, University of Sheffield,\\
       Sheffield S3 7RH, England
       E-mail: \email {L.Roszkowski@sheffield.ac.uk}}

\author{Roberto Ruiz de Austri\\
       Instituto de F\'isica Corpuscular, IFIC-UV/CSIC \\
       Valencia, Spain \\
       E-mail: \email{rruiz@ific.uv.es}}

\author{Roberto Trotta\\
       Astrophysics Group, Imperial College London \\
       Blackett Laboratory, Prince Consort Road, London SW7 2AZ, UK\\
       E-mail: \email{r.trotta@imperial.ac.uk}}

\abstract{ We employ two different statistical tests to examine
  whether, in the framework of the Constrained MSSM, the
  experimentally determined values of $\brbsgamma$ and the anomalous
  magnetic moment of the muon, $\gmtwo$ are consistent with each
  other. Our tests are designed to compare the theoretical predictions
  of the CMSSM in data space with the actual measurements, once all of
  the CMSSM free parameters have been integrated out and constrained
  using all other available data.  We investigate the value of
  $\gmtwo$ as obtained by using $\epem$ data alone (which shows a
  $\sim 3\sigma$ discrepancy with the Standard Model prediction) and
  as obtained based on $\tau$ decay data (which shows a much milder,
  $\sim 1\sigma$ discrepancy). We find that one of our tests returns
  either a statistically inconclusive result or shows weak evidence of
  tension between $\brbsgamma$ and the $\epem$--data based value of
  $\gmtwo$. On the other hand, our second test, which is more
  stringent in this application, reveals that the joint observations
  of $\brbsgamma$ and $\gmtwo$ from $\epem$ data alone are
  incompatible within the CMSSM at the $\sim 2\sigma$ level. On the
  other hand, for both tests we find no significant tension between
  $\brbsgamma$ and the value of $\gmtwo$ evaluated using $\tau$ decay
  data. These results are only weakly dependent on the three different
  priors that we employ in the analysis. We conclude that, if the
  discrepancy between the Standard Model and the experimental
  determinations of $\gmtwo$ is confirmed at the $\sim 3\sigma$ level,
  this could be interpreted as strong evidence against the CMSSM. }


\begin{document}

\section{Introduction}\label{sec:intro}

Softly broken low-energy supersymmetry (SUSY) is considered to be perhaps 
the most promising theory beyond the Standard Model (SM). Not only does it
provides an elegant solution to the hierarchy problem~\cite{Martin:1997ns} 
but also naturally accommodates gauge coupling unification~\cite{Ellis:1990wk} 
and offers a clue to the dark matter (DM) problem in the 
Universe~\cite{Komatsu:2008hk}.

On the other hand, without specifying a complete underlying mechanism
of SUSY breaking, the general Minimal Supersymmetric Standard Model
(MSSM) suffers from a large number of SUSY-breaking soft parameters
which are poorly determined. Motivated by a natural link between SUSY
and grand unified theories (GUTs), over the last several years it has
become customary to impose various boundary conditions at the GUT
scale and explore resulting SUSY phenomenology. The most popular model
of this class is the Constrained MSSM (CMSSM)~\cite{kkrw94}, includes the 
minimal supergravity model (mSUGRA)~\cite{sugra}. In this scheme one 
defines all SUSY parameters at the unification scale $\mgut $ and next 
employs the Renormalization Group Equations (RGEs) to evolve them down and 
compute the couplings and masses in an effective theory valid at the
electroweak scale. The CMSSM is defined in terms of four continuous free
parameters: common scalar ($\mzero$), gaugino ($\mhalf$) and
tri-linear ($\azero$) mass parameters (all specified at the GUT
scale), plus the ratio of Higgs vacuum expectation values $\tanb$; and
one discrete parameter $\sgn(\mu)$, where $\mu$ is the Higgs/higgsino mass
parameter whose square is computed from the conditions of
radiative electroweak symmetry breaking (EWSB).

The phenomenology of the CMSSM has been studied in a vast number of
papers. The usual approach has been to explore the
model by performing fixed grid scans in $\mhalf$ and $\mzero$ for
fixed, ``representative'' values of $\tanb$ and for
$\azero=0$~\cite{grid-cmssm}, and also
for fixed values of SM parameters, \eg, the top mass $\mtop$, which
can have a large impact on results especially at large $\mzero$. Also,
the model's predictions for observable quantities, \eg, the relic
abundance $\abundchi$ of the neutralino, or Higgs and superpartner
masses have been compared with experimental data in a simplified way:
if the predicted values is within some arbitrary range, typically
$1\,\sigma$ or 90\%~CL then the point is treated as ``allowed'';
otherwise it is rejected. Theoretical errors are also typically
ignored. A $\chisq$ approach applied in~\cite{ehow02,ehow06,ehhow07} addressed
the latter problems. On the other hand, in those papers it was
advocated to reduce the effective number of CMSSM parameters by using
the well-measured value of $\abundchi$ to determine a ``surface''
in the model's parameter space was somewhat questionable as its shape
and ``thickness'' can critically depend on the actual value of $\mtop$,
especially at large $\mzero$ (compare fig.~4 in~\cite{rrt3}) and also by including a
``fudge factor'' in the definition of $\chisq$ in order to suppress
the contribution of the large  $\mzero$ region (see eq.~(1)
in~\cite{ehow06}). In a more recent analysis~\cite{buchellislatest}
that element of the $\chisq$ analysis has been abandoned. On the
other hand, the conclusions  of~\cite{buchellislatest} heavily rely on
the somewhat uncertain discrepancy between the SM and the experimental
determinations of the anomalous magnetic moment of the muon $\gmtwo$. 

Over the last few years a new approach based on Bayesian statistics
linked with either Markov Monte Carlo Chain
(MCMC)~\cite{bg04,al05,allanach06,rtr1,alw06,rrt2,rrt3,Trotta:2006ew,rrts,Allanach:2008tu,Allanach:2008iq,Roszkowski:2009sm} 
or Nested Sampling (NS) scanning methods~\cite{Feroz:2007kg,BenModelComp,tfhrr1}
has been successfully applied in a well defined statistical
framework~(see, \eg,~\cite{Trotta:2008qt}).  Furthermore the priors
issue, considered as a ``soft spot'' of the Bayesian approach, has
been recently thoroughly addressed and has been shown to embody in a
quantitative manner the physical fine-tuning of the theory,
see~\cite{ccr08}.

One of the outcomes of the most recent and more sophisticated scans
has been to realize that even the CMSSM, with its relative economy of
free parameters, remains presently somewhat underconstrained by
currently available data, thus leaving large regions of CMSSM
parameters allowed~\cite{tfhrr1}. Despite this, it was pointed out
in~\cite{rrt3}, and investigated in more detail in~\cite{tfhrr1}, that
there appears to exist a certain ``tension'' between the current
measurements of $\brbsgamma$ (hereafter denoted by $\bsg$ for brevity)
and $\gmtwo$, in the sense that the two observables favor different regions of the CMSSM parameters space (compare
figs.~8 and 10 in~\cite{tfhrr1}). This is because the $\brbsgamma$
constraint favors the focus point (FP)
region~\cite{Chan:1997bi,focuspoint-fmm}, as the (always positive) 
charged Higgs/top contribution has to be large enough so that,
starting from the SM central value of $3.12\times10^{-4}$, the
(negative, for $\mu>0$) chargino/stop contribution can bring the sum
down to the experimental central value of $3.55\times10^{-4}$. This requires
the charged Higgs to be light enough and the stop (or chargino, or both) to be
heavy enough. Both conditions are satisfied in the FP region. On the
other hand, large corrections to the $\gmtwo$ values arise mostly in
the low--mass region. We feel that it would be interesting 
to further investigate this tension, and to develop statistical tools
to quantify the possible incompatibility of the two observations
within the theoretical model. A strong tension between
the data would then be interpreted either as a sign of an undetected
(or underestimated) systematic error in one of the data sets, or as a
sign that the theoretical model is at odds with the data and hence is
disfavored. One first evaluation of the tension has been carried out
in~\cite{BenModelComp} using a model comparison test, returning
however an inconclusive result. The purpose of this paper is to
re-consider the problem of the tension between these two observables
by addressing it with a novel statistical test, called ``the
predictive likelihood ratio test'', or the $\ltest$ introduced
below.  We clarify that the reason why the test based on model
comparison performed in~\cite{BenModelComp} is inconclusive can be
traced back to the orientation of degeneracies in data space, a
feature that in this particular context makes the model comparison
test less stringent than the new test introduced here.

The paper is organized as follows. In section~\ref{sec:theory} we
introduce the statistical framework and define the new test for the
compatibility of observables based on the predictive data distribution
(the $\ltest$) as well as present the test based on the model comparison (the $\rtest$). In sec.~\ref{sec:cmssmtests} we specify the
theoretical model and its parameters and the priors we consider and we apply the statistical tests to the
CMSSM. We present our numerical results in
section~\ref{sec:results}. Our conclusions are given in
section~\ref{sec:concl}.

\section{Setup}
\label{sec:theory}
\subsection{Statistical framework}
\label{sec:stats}

We follow here the notation and conventions of our previous
works~\cite{rtr1, rrt2, tfhrr1}. We denote the set of parameters of
the model $\mdl$ under consideration (here the CMSSM) by $\params$, and by
$\nuis$ all other relevant parameters, the so-called {\em nuisance
parameters}, which here include relevant SM quantities. Both sets form
our {\em basis  parameters} 
\be
\basis = (\params,\nuis).
\label{basis:eq}
\ee
Bayesian inference is based on Bayes' theorem which reads
\be \label{eq:bayes}
p(\basis | \data, \mdl) = \frac{p(\data |
\basis, \mdl) p(\basis | \mdl)}{p(\data|\mdl)}. \ee
The quantity $p(\basis | \data, \mdl)$ on the l.h.s.~of eq.~\eqref{eq:bayes}
is called a {\em posterior probability density function} (posterior
pdf, or simply a
{\em posterior}).  On the r.h.s., the quantity $p(\data |\basis, \mdl)$,
taken as a function of $\basis$ for {\em fixed data} $\data$, is
called the {\em likelihood}.  The likelihood supplies the information provided by
the data.  The quantity
$p(\basis|\mdl)$ denotes a {\em prior probability density function}
(prior pdf, or simply a {\em prior}) which encodes our state of knowledge about the
values of the parameters in $\basis$ before we see the data. The prior state
of knowledge is then updated to the posterior via the likelihood.

Finally, the quantity in the denominator is called {\em the evidence} or
{\em model likelihood}, which is obtained by computing the average of
the likelihood under the prior (so that the r.h.s. of
eq.~\eqref{eq:bayes} is properly normalized to unity probability), 
\be \label{eq:evidence}
p(\data|\mdl) = \int  p(\data | \basis,\mdl) p(\basis|\mdl) \dr\basis . 
\ee
If one is interested in constraining the model's parameters, the
evidence is merely a normalization constant, independent of $\basis$,
and can therefore be dropped. However, the evidence is very useful in
the context of Bayesian model comparison (see
e.g.~\cite{Trotta:2005ar,stv08} and~\cite{alw06,BenModelComp} for
recent applications to the CMSSM). One of the main goals of this paper
is to develop and apply to the CMSSM a new evidence-based statistical
test on the consistency of two or more observables within a given
theoretical model.
Taking into account that one might wish to consider different models,
all of the above relations have been conditioned explicitly on the
model under consideration, $\mdl$. However, in the following we will
drop the explicit conditioning on $\mdl$ since we only work in the
framework of a single, given model in this paper, namely the CMSSM.
 
\subsection{The predictive likelihood ratio test \boldmath ($\ltest$)} 

 Let us split the full data set of $n$ observables $\data$ (which will
 be given below) as $\data =
\{ \test, \dass\}$. Suppose and that we are interested in testing the
compatibility of the observations within a subset
$\test=\{\test_1,\dots,\test_k\}$, $k<n$, conditional on the observed
values for the second part of the data set, $\dass = \{ \data_{k+1},
\dots,\data_n \}$, which are considered as external (independent)
 constraints which are assumed to be correct.  We are thus interested
 in evaluating the {\em 
conditional evidence} $p(\test | \dass )$, which represents the
probability of measuring data $\test$ given that data $\dass$ have
been gathered for the remaining $n-k$ observables. In other words,
this conditional probability can be interpreted as the predictive
probability for a measurement of the observables $\test$ given what
has been observed for the other quantities. As a consequence of the
basic probability manipulation rules, the conditional evidence can be
written as \be
\label{eq:jointev} p(\test|\dass) = \frac{p(\test, \dass)}{p(\dass)} ,
\ee (recall that we are dropping the conditioning on the model $\mdl$
which is understood). On the r.h.s., the joint evidence $ p(\test,
\dass) $ is the probability of measuring the joint data set within the
assumed model, {\em independently of the actual true values of the
model's parameters $\basis$,} which have been integrated out in the
computation of the evidence, see eq.~\eqref{eq:evidence}. The joint
evidence has to be evaluated {\em as a function of the possible
outcomes of the observations of the data set $\test$}. This requires
evaluating the evidence for a series of possible values for $\test$,
at each time integrating over the full parameter space of the
model. The possible data realizations $\test$ are different outcomes
for the measurements (e.g., different means) given the experimental
noise, i.e., the reported error of the central value. At the same
time, the data set $\dass$ are held fixed at their actual observed
values, for, as stated above, we assume that this part of the data
set is trustworthy and can be used to constrain the model's
parameters. Notice that while the central values of the
data set  $\dass$ are assumed to be correct, the uncertainty on their
value is automatically fully accounted for, since we integrate over
all the model's parameters when computing the evidence and we include both
experimental and theoretical errors on $\dass$. 

Once $p(\test|\dass)$ is obtained as a function of $\test$, its
evaluation at the observed value $\test = \test^\text{obs}$ allows one
to determine the compatibility of the observed data realization
$\test^\text{obs}$ with the model and the rest of the data, by
evaluating the relative probability of obtaining such a realization
compared to the maximum probability for the data set in question. Let
us denote by $\test^\text{max}$ the values of the data that maximises $p(\test
| \dass)$. Then the relevant quantity to consider is the ratio
\be \label{eq:ltest}
\predlike(\test^\text{obs}|\dass) \equiv  \frac{p(\test^\text{obs} |
  \dass)}{p(\test^\text{max} | \dass)}  = \frac{p(\test^\text{obs},
  \dass)}{p(\test^\text{max} , \dass)},  \qquad\qquad\qquad (\ltest), 
\ee
where we have used eq.~\eqref{eq:jointev} in the second equality. This
is analogous to a likelihood ratio in data space, {\em but integrated
over all possible values of the parameters of the model}. We call the
$\ltest$ the {\em predictive likelihood ratio test}. If
$\predlike(\test^\text{obs}|\dass) \sim 1$, this shows that both data
sets are compatible with each other and with the model's assumptions
(including the prior choice), and therefore we can legitimately use
them together to constrain the parameters of the model $\mdl$. If
however $\predlike(\test^\text{obs}|\dass) \ll 1$, we should doubt the
consistency of the data $\test$ (perhaps considering the possibility
of systematic effects) or the model's assumptions (i.e., the choice of
model or of the assumed form and/or ranges of its
priors). If $\ltest$ comes out to be weakly dependent
on the prior, then this will give us more confidence that the
conclusions of the statistical test when applied to the assumed model are
robust.

A simple example of the application of the $\ltest$ method to a toy
linear model is presented in Appendix~\ref{appendix}.

\subsection{The model comparison test \boldmath ($\rtest$)} \label{sec:rtest}

A different compatibility test has been employed
by~\cite{BenModelComp}, following earlier applications in
cosmology~\cite{consistency}. The gist of what was called 
``model comparison test'' there can be summarized as follows
(see~\cite{BenModelComp} for full details).

The idea is to perform a Bayesian model comparison test between two
hypotheses, namely $\Hy_0$, stating that the data $\test$ under
scrutiny are all compatible with each other and with the model, versus
$\Hy_1$, purporting that the observables are incompatible (within
the assumed model) and hence
tend to pull the constraints in different regions of parameter
space. For $k>1$, the Bayes factor between the two
hypotheses, giving the relative probabilities (odds) between $\Hy_0$
and $\Hy_1$ is given by 
\be \label{eq:R1}
\Rr = \frac{p(\test | \dass, \Hy_0)}{\prod_{i=1}^k  p(\test_i | \dass, \Hy_1)}.    
\ee
Writing again the conditional evidences in terms of the joint
evidences, e.g. $p(\test | \dass, \Hy_0) = p(\test, \dass |
\Hy_0)/p(\dass|\Hy_0)$, and noting that $p(\dass|\Hy_0) =
p(\dass|\Hy_1)$ (which follows because the evidence from the data we
are not testing does not depend on the hypothesis being considered), 
eq.~\eqref{eq:R1} can be recast as
\be \label{eq:rtest}
\Rr(\test) = \frac{p(\test,\dass | \Hy_0)}{\prod_{i=1}^k   p(\test_i,
  \dass | \Hy_1)}  p(\dass |\Hy_0)^{k-1}  \qquad\qquad\qquad (\rtest, k>1).     
\ee
If instead $k=1$, i.e., $\test = \test_1$ and we wish to test the
consistency of one single new observation, then eq.~\eqref{eq:rtest}
needs to be modified to 
\be \label{eq:rtestk1} \Rr(\test_1) = \frac{p(\test_1, \dass |
\Hy_0)}{p(\test_1 | \Hy_1)p(\dass |\Hy_1)} \qquad\qquad\qquad (\rtest,
k=1).  \ee Eqs.~\eqref{eq:rtest} and \eqref{eq:rtestk1} are then
evaluated at the observed value of the data sets being tested, i.e.~
for $\test = \test^\text{obs}$. If $\ln \Rr( \test^\text{obs}) > 0$,
this is evidence in favour of the hypothesis $\Hy_0$ that the data are
compatible. If instead $\ln\Rr( \test^\text{obs}) < 0$ the alternative
hypothesis $\Hy_1$ that there is a tension among the data (and the
model) is preferred. More quantitatively, the strength of evidence for
either case can be assessed against so-called ``Jeffreys' scale'',
which we report in table~\ref{tab:Jeff} along with our (slightly
modified) convention for denoting the different levels of evidence.

\begin{table}[tbh!]
\centering
 \begin{tabular}{|l | l | l|} 
 \hline
  $|\ln \Rr |$ & Odds & Strength of evidence \\
 \hline
 $<1.0$ & $\lsim 3:1$  & Inconclusive \\
 $1.0$ & $\sim 3:1$ & Weak evidence \\
 $2.5$ & $\sim 12:1$  & Moderate evidence \\
 $5.0$ & $\sim 150:1$ & Strong evidence \\
 \hline
\end{tabular}
\caption{Empirical scale for evaluating the strength of evidence (so-called
``Jeffreys' scale''). Threshold values are empirically set, and
they occur for values of the logarithm of the Bayes factor between the hypotheses of
$|\ln \Rr|=1.0$, 2.5 and 5.0. The right-most column gives our
convention for denoting the different levels of evidence above
these thresholds, according to the prescription in~\cite{gt07}.\label{tab:Jeff} }
\end{table}

In applying the test to the CMSSM below we will consider the cases
$k=1$ and $k=2$ with, as mentioned above, the two pieces of data being
tested for mutual consistency being $\bsg$ and $\deltagmtwo$ (the
latter both from $\tau$ decay and $e^+e^-$ data separetely).

\section{An application to the CMSSM}
\label{sec:cmssmtests}

Before  we apply the above formalism to the CMSSM, we first specify
the priors tested and the experimental constraints.

\subsection{Choice of priors and data}
\label{sec:data}

In order to assess the robustness of our results with respect to
plausible changes of priors, we consider three different classes of
priors:
\begin{itemize}
\item {\bf flat prior:} flat on $\mzero, \mhalf, \azero, \tanb$, with
  ranges as given in section 3.2 of~\cite{tfhrr1}; 

\item {\bf log prior:} flat on $\ln\mzero, \ln\mhalf, \azero, \tanb$,
  with ranges as given in section 3.2 of~\cite{tfhrr1}; 

\item {\bf CCR mSUGRA prior:} flat on $\mzero, \mhalf, \azero, B$
  but with an effective  ``penalty term'' that naturally leads to low
  fine tuning among SUSY parameters.
\end{itemize}

Unlike the first two priors, which refer to the CMSSM parameterization
in terms of its parameters $\theta = (\mhalf, \mzero,\azero, \tanb)$,
the third prior, as introduced in ref.~\cite{ccr08} by Cabrera, Casas and Ruiz de Austri (hence the name), is applied to mSUGRA with its basic
parameters $\mhalf, \mzero,\azero, B$, augmented by the top Yukawa
coupling $y_t$. A marginalization over $\mu$ selects the value $\mu_0$
that reproduces the experimental value of $M_Z$.  As shown in
ref.~\cite{ccr08}, it is also convenient and natural to trade the
parameter $B$ for $\tanb$ and $y_t$ for the top mass $m_t\propto
y_t\sin\beta $.  This procedure results in an effective prior
\begin{equation}
p_\text{eff}(m_t, \mzero, \mhalf, \azero, \tanb) =
J|_{\mu=\mu_0}  p(y_t, \mzero, \mhalf, \azero, B, \mu=\mu_0).
\end{equation} 
Assuming a flat prior on the parameters $\mhalf, \mzero, \azero, B$
and a log prior on $y_t$, the Jacobian term acts as an effective
``penalty term'' that favors lower values of $\mu$ and $\tanb$, and
thus leads to less fine-tuning as in the focus point
region. In the CMSSM this
corresponds to large $\mzero$. On the other hand, the changing of
parameters from $B$ to $\tanb$ favors large $\mhalf$ because of the
$B$ dependence on $\mhalf$ in the RGEs.  (See ref.~\cite{ccr08} for
more details.)

As we shall see, our results are largely insensitive to the
choice of priors, which indicates a remarkable robustness of this
statistical test. This can be traced backed to the fact that the
parameters within the model are fully integrated out in the
computation of the predictive probability.

\begin{table}[b!th]
\centering
\begin{tabular}{|l | l l l | l|}
\hline
Observable &   Mean value & \multicolumn{2}{c|}{Uncertainties} & ref. \\
&   $\mu$      & ${\sigma}$ (exper.)  & $\tau$ (theor.) &
\\ \hline
$\deltaamususy \times 10^{10}$       &  29.5 & 8.8 &  1.0 &
\cite{gm2} ($\epem$ data)\\ 
&  8.9 & 9.5 &  1.0 & \cite{gm2-tau} ($\tau$ data) \\\hline
$\brbsgamma \times 10^{4}$ &
3.55 & 0.26 & 0.21 & \cite{hfag}   \\
\hline
\end{tabular}
\caption{Summary of the observables $\test$ being tested for
  consistency. \label{tab:datatest}} 
\end{table}

The focus of this paper is to test for consistency the measured values
of $\bsgamma$ and the anomalous magnetic moment of the muon, $\gmtwo$.
For the latter, we consider two sets of measurements: the first is
based on $e^+e^-$ data, and it gives a $\sim3.2\sigma$ discrepancy with
the SM predicted value~\cite{gm2}; the second one employs
$\tau$ decay data to evaluate the SM hadronic contribution to $\gmtwo$
instead, which leads to a much better agreement,
$\deltaamususy=(8.9\pm 9.5)\times 10^{-10}$~\cite{gm2-tau}. These values and their
uncertainties are listed in the top part of Table~\ref{tab:datatest}.

As regards $\brbsgamma$, for the new SM prediction
we obtain the value of $(3.12\pm 0.21)\times10^{-4}$.\footnote{The
value of $(3.15\pm 0.23)\times10^{-4}$ originally derived in
ref.~\cite{ms06-bsg,mm-prl06} was obtained for slightly different
values of $\mtpole$ and $\alphas$. Note that, in treating the error
bar we have explicitly taken into account the dependence on $\mtpole$
and $\alphas$, which in our approach are treated parametrically. This
has led to a slight reduction of its value.}  We compute SUSY contribution to
$\brbsgamma$ following the procedure outlined in
refs.~\cite{dgg00,gm01} which was extended in
refs.~\cite{or1+2,for1+2} to the case of general flavor mixing. In
addition to full leading order corrections, we include large
$\tanb$-enhanced terms arising from corrections coming from beyond the
leading order and further include (subdominant) electroweak
corrections. 

All the other experimental values of the collider and cosmological observables
that we assume in order to perform the compatibility test for
$\deltaamususy$ and $\brbsgamma$ are listed in
table~\ref{tab:meas}. We refer to~\cite{rtr1,rrt2,tfhrr1} for details
about the computation of each quantity and for justification of the
theoretical errors adopted, as well for a detailed description of the
likelihood function. 
In particular, points that do not fulfil the conditions of radiative
EWSB and/or give non-physical (tachyonic) solutions are
discarded. Also, we take $\mu>0$, because of its correlation with sign
of $\deltagmtwo$.

\begin{table}
\centering
\begin{tabular}{|l | l l l | l|}
\hline
Observable &   Mean value & \multicolumn{2}{c|}{Uncertainties} & ref. \\
&   $\mu$      & ${\sigma}$ (exper.)  & $\tau$ (theor.) &
 \\\hline
\multicolumn{5}{|c|}{Nuisance paramaters}  
\\ \hline
$\mtpole$           &  172.6 GeV    & 1.4 GeV&  N/A & \cite{topmass:mar08} \\
$m_b (m_b)^{\overline{MS}}$ &4.20 GeV  & 0.07 GeV & N/A &  \cite{pdg07} \\
$\alphas$       &   0.1176   & 0.002 &  N/A & \cite{pdg07}\\
$1/\alphaemmz$  & 127.955 & 0.03 &  N/A & \cite{Hagiwara:2006jt} \\ \hline
\multicolumn{5}{|c|}{Observables (measured)}  
\\ \hline
$M_W$     &  $80.398\gev$   & $25\mev$ & $15\mev$ & \cite{lepwwg} \\
$\sineff{}$    &  $0.23153$      & $16\times10^{-5}$
               & $15\times10^{-5}$ &  \cite{lepwwg}  \\
$\delmbs$     &  $17.77\ps^{-1}$  & $0.12\ps^{-1}$  & $2.4\ps^{-1}$
& \cite{cdf-deltambs} \\
$\brbtaunu \times 10^{4}$ &  $1.32$  & $0.49$  & $0.38$
& \cite{hfag} \\
$\abundchi$ &  0.1099 & 0.0062 & $0.1\,\abundchi$& \cite{wmap5yr} 
 \\\hline
\multicolumn{5}{|c|}{Observables (limits)}  \\
  &  Limit (95\%~\cl)  & \multicolumn{2}{r|}{$\tau$ (theor.)} & ref. \\\hline\hline  
$\brbsmumu$ &  $ <5.8\times 10^{-8}$
& \multicolumn{2}{r|}{14\%}  & \cite{cdf-bsmumu}\\
$\mhl$  & $>114.4\gev$\ (SM-like Higgs)  & \multicolumn{2}{r|}{$3 \gev$}
& \cite{lhwg} \\
$\zetah^2$
& $f(m_h)$\ (see ref.~\cite{rtr1})  & \multicolumn{2}{r|}{negligible}  & \cite{lhwg} \\
$m_{\tilde{q}}$ & $>375$ GeV  & & 5\% & \cite{pdg07}\\
$m_{\tilde{g}}$ & $>289$ GeV  & & 5\% & \cite{pdg07}\\
other sparticle masses  &  \multicolumn{3}{c|}{As in table~4 of
 ref.~\cite{rtr1}.}  & \\ \hline 
\end{tabular}
\caption{Summary of the observables $\dass$ used in the analysis, on
  which the consistency test is conditional. Upper part: measurements
  on nuisance (SM) parameters. (N/A stands for ``not
  applicable''.) Central part: Observables for which a 
  positive measurement has been
made. Lower part: Observables for which only limits currently
exist.  For details, see the treatment in ref.~\protect\cite{rtr1,rrt2, tfhrr1}.
\label{tab:meas}}
\end{table}

\subsection{Applying the $\ltest$ to the CMSSM}\label{sec:applcmssm}

We are interested in assessing the compatibility of $\test = \{ \bsg,
\deltagmtwo \}$, while assuming all the other data (denoted by $\dass$) to be
believable. We remind the reader at this point that we are concerned
with making predictions in data space, and {\em not} in parameter
space, as it is usually done. We are not interested in constraining the
parameters of the model here, but instead integrate over all
their possible values. Therefore, the resulting values of $\bsg$ and $\deltagmtwo$
should be understood to represent the mean values that are predicted
to be obtained experimentally for the respective quantities within the
CMSSM, once all the other constraints on the r.h.s. of the conditioning
bar ($\dass$, as in table~\ref{tab:meas}) are taken into account
(including their experimental and theoretical uncertainties). 

We thus evaluate the evidence and compute the predictive probability
on a grid of values for $(\bsg, \deltagmtwo)$ representing the possible
outcomes for the central value of the observation. At each point we
keep the same experimental error  as the one that has been effectively reported by the
experiments (adding the theoretical error on top), as given in
table~\ref{tab:datatest}. In other words, we consider different
possible outcomes for the central values but with fixed instrumental noise
properties, which is a reasonable assumption.

The computation of the evidence is numerically costly, as it involves
an 8--dimensional integral over the whole parameter space for every
choice of data values one wishes to test. We employ a modified version
of the \texttt{SuperBayeS} package~\cite{rrt2} including the MultiNest
algorithm~\cite{tfhrr1}, which allows one to compute the evidence and from
there, the predictive probabilities involved in the $\ltest$. Despite
MultiNest's high efficiency, each evidence evaluation still requires
about 3 days of CPU time on 4 3.00 GHz Intel Woodcrest processors. Appendix
of~ref.~\cite{tfhrr1} provides a full description of how the uncertainty on the
value of the evidence is evaluated with MultiNest. This uncertainty is
then propagated to the uncertainty on the $\ltest$ of
eq.~\eqref{eq:ltest}. 

We scan over the following central values for
the experimental outcomes, chosen to bracket the actually observed
values:
\begin{align}
\brbsgamma \times 10^4: &\quad  1.5,\ldots, 4.0~~{\rm in~ intervals~ of~ 0.5}\\
\deltagmtwo \times 10^{10}: & \quad 0,\ldots, 40~~{\rm in~ intervals~ of~ 5}.
\end{align}
As for the experimental noise, we fix this to the actually reported
value for the real observation, supplemented by a suitable theoretical
error, as given in table~\ref{tab:datatest}. When considering the two
different experimental determinations of $\deltagmtwo$ (one based on
$\tau$ decay data and one based on $e^+e^-$ data alone), we should in
principle repeat our test using the reported experimental error for
each of the observations. However, the reported experimental errors on
$\deltagmtwo$ for the two determinations of the quantity are very
similar (within about 10\%) and therefore we employ the uncertainty
reported in using the $\epem$ data for both. This approximation is not
expected to influence significantly our result.

\section{Numerical results}
\label{sec:results}

It is interesting to consider both the $\ltest$ and the $\rtest$, for
each of them is sensitive to possible tensions between the observables
in a different way and may in general give different results. (This is
demonstrated in a toy model example in Appendix~\ref{appendix}.) This
is in fact not surprising, for while different measurements can be
compatible with each other and also compatible with the model being
fitted in only one way if all the measurements are correct and the
theory is the right model, there are many different ways in which an
incompatibility could manifest itself. The $\ltest$ asks what is the
probability of measuring a certain value for the data subset $\test$
(relative to the maximum probability achievable under the model) given
what is known about the model from the remaining data $\dass$. The
$\rtest$ instead tries to enforce consistency between the data being
tested and the remaining data sets, by looking for values of $\test$
that are jointly compatible with the parameter space singled out by
$\dass$. These two approaches show subtle differencies and in general
play out differently whenever a genuine tension between the
observables exists.
 
Furthermore, the two tests are evaluated on different scales: the
$\ltest$ being of the form of a likelihood ratio test can be evaluated
on a significance scale analogous to the usual $\Delta \chi^2$ rule,
while the $\rtest$ (representing odds between two hypotheses) should
be assessed against Jeffreys' scale for the strength of
evidence. Another issue to consider is that in general the two tests
favour different degenerate regions of data space (see
fig.~\ref{fig:toys} in Appendix~\ref{appendix} for an illustration)
and one can easily imagine situations where one of these regions is
more constrained than the other, due to the structure of the model. In
this case, the test that exhibits more power along this more
constrained degenerate region will appear to be more stringent.

\subsection{Results for the $\ltest$}

\begin{figure}[tbh!]
\begin{center}
\includegraphics[width=\ww]{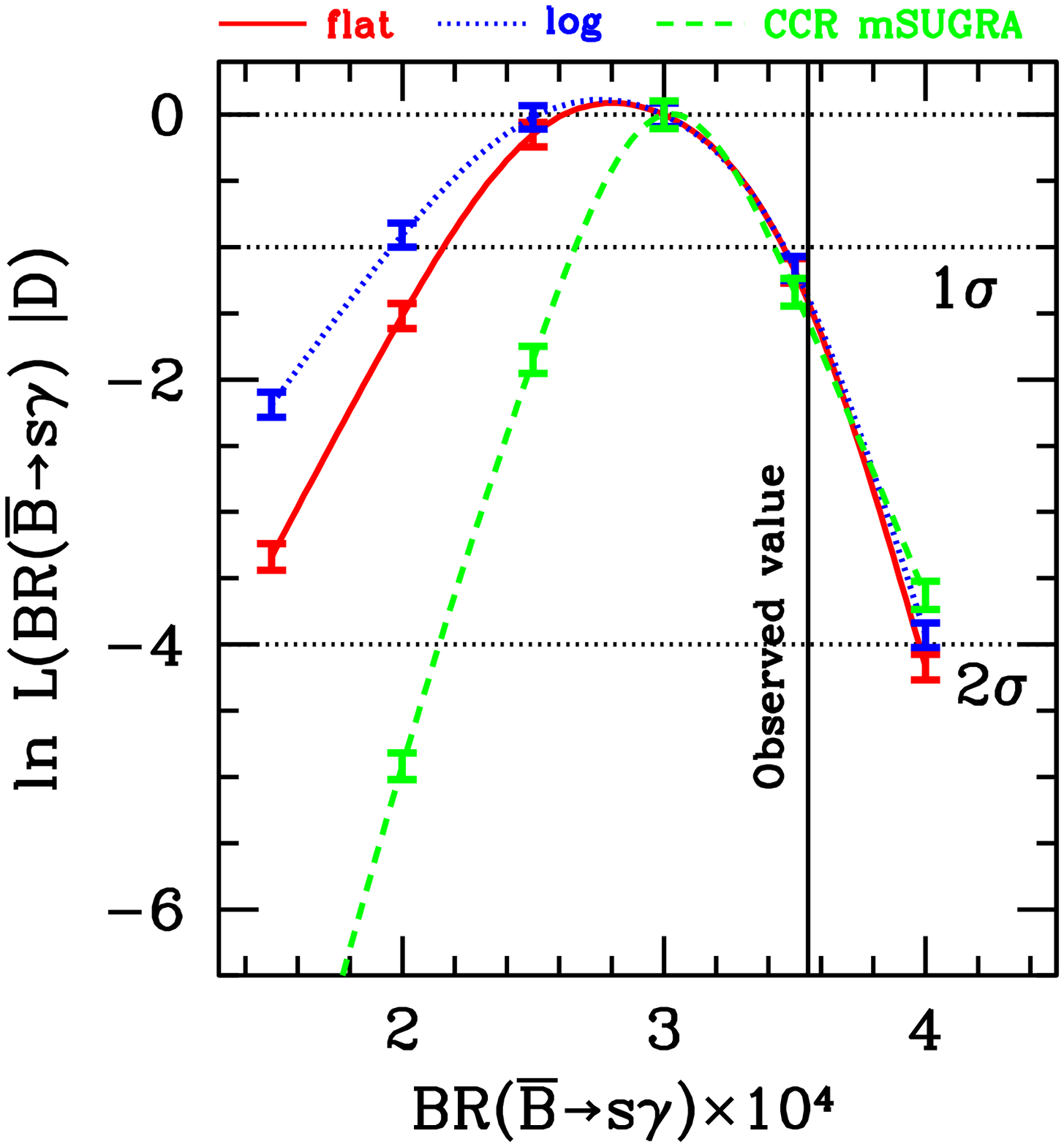}
\includegraphics[width=\ww]{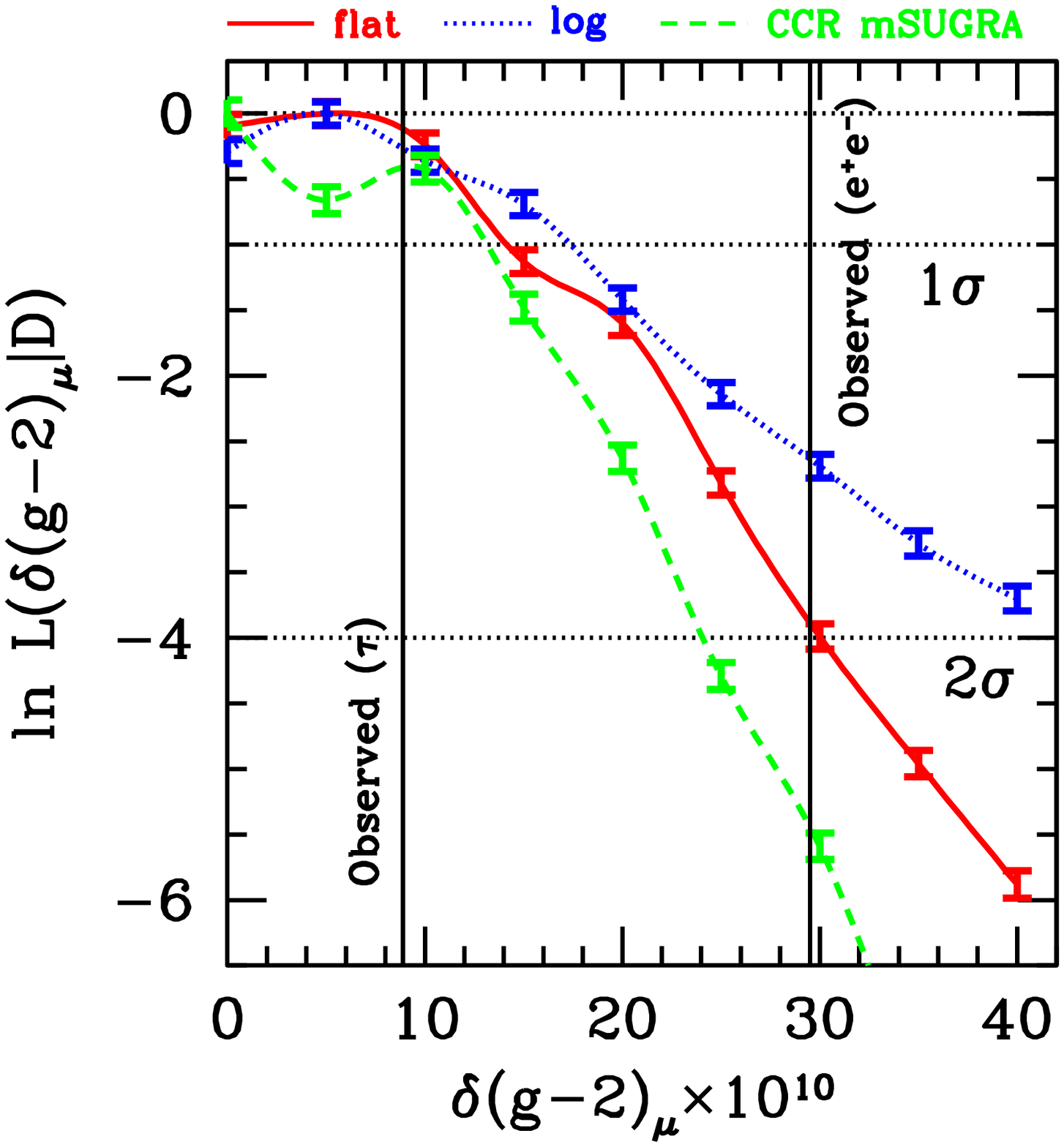} \\ 
\caption[test]{Predictive data distribution ($\ltest$) for $\bsg$
  (left panel) and $\deltagmtwo$ (right panel) in the CMSSM for three
  different choices of priors: flat prior (red/solid), log prior
  (blue/dotted) and the CCR mSUGRA prior (green/dashed). The predictive
  distributions are conditional on all other observations, excluding
  $\deltagmtwo$ and $\bsg$. The vertical lines give the actual
  measured values. The errorbars denote the location at which the
  predictive probability has been computed (and its error), while the
  lines are a smoothed spline.} 
\label{fig:1D_predictive} 
\end{center}
\end{figure}

We begin by employing the $\ltest$ to separately test the consistency
between $\test_1 = \bsg$ and the other observables $\dass$ (but
excluding from the latter $\deltagmtwo$) and between $\test_1 =
\deltagmtwo$ and the other observables (but excluding from the latter
$\bsg$). Notice in particular that we do include the dark matter
constraint in the assumed data $\dass$. The outcome of these two tests
is shown in fig.~\ref{fig:1D_predictive} and reported in
table~\ref{tab:ltest1D}. In the left panel of
fig.~\ref{fig:1D_predictive}, we plot the quantity
$\predlike(\brbsgamma| D)$ as a function of the possible outcome of
the experimental observation, with the actual observed central
value indicated by a vertical, solid line. In the right panel of
fig.~\ref{fig:1D_predictive}, we plot instead $\predlike(\deltagmtwo|
D)$ as a function of the possible measured values of $\deltagmtwo$,
indicating by vertical lines the actual observed values from $\epem$
and $\tau$ data.

The CMSSM, once all the observations other than
$\deltagmtwo$ are accounted for, tends to predict a $\bsg$ value close
to the SM prediction, $\brbsgamma \simeq 3.12 \times 10^4$ (the precise
value depending on the actual values of SM input parameters,
especially $\mtop$ and $\alphas$),  with predominantly small
negative corrections arising from chargino-stop loop contributions. This is shown by
the peak in the predictive distribution, which occurs at around $\brbsgamma
\sim 3 \times 10^4$. The experimental central value $ 3.55 \times 10^4$ is
within about 1$\sigma$ of the most likely value, thus it is not
significantly in tension with the other observables (see top part of
table~\ref{tab:ltest1D}).  

\begin{table}
\centering
 \begin{tabular}{|l |  l | l |} 
  \hline
 Prior & $\ln \predlike(\test_1 |D)$ & Interpretation  \\\hline\hline
  \multicolumn{3}{|l| }{$\test_1 = \brbsgamma$} \\\hline
 Flat &  $ -1.63 \pm 0.11 $ &  Not significant  ($1.28\sigma$) \\
 Log & $  -1.43 \pm 0.12 $ &  Not significant ($1.20\sigma$)\\
  CCR mSUGRA &  $-0.89 \pm 0.13$ & Not significant ($<1\sigma$)  \\\hline
 \multicolumn{3}{|l|}{$\test_1 = \deltagmtwo$ from $\epem$ data} \\\hline
 Flat &  $ -3.99 \pm 0.10 $ & Incompatible at 95.4\% significance   \\
 Log & $  -2.69 \pm 0.10 $ & Not significant ($1.64\sigma$) \\
  CCR mSUGRA &  $-5.59  \pm 0.10$ & Incompatible at  98.2\% significance \\\hline
 \multicolumn{3}{|l|}{$\test_1 = \deltagmtwo$ from $\tau$ decay data} \\\hline
 Flat &  $-0.24\pm 0.10 $ & Not significant  ($<1\sigma$)   \\
 Log & $  -0.38 \pm 0.08 $ & Not significant  ($<1\sigma$)   \\
  CCR mSUGRA &  $-0.30 \pm 0.08$ & Not significant  ($<1\sigma$) \\
  \hline
\end{tabular}
\caption{Results of the $\ltest$, testing for consistency of
  $\brbsgamma$ with all other data (excluding $\deltagmtwo$) and for
  consistency of $\deltagmtwo$ with all other data (excluding
  $\brbsgamma$).\label{tab:ltest1D} }
\end{table}

Turning next to the anomalous magnetic moment of the muon (right panel
of fig.~\ref{fig:1D_predictive}), the
predictive probability is largest for $\deltagmtwo \sim 0$, as might
be expected from noting that only a small fraction of the CMSSM
parameter space gives rise to sizable SUSY corrections to
$\deltagmtwo$. The probability remains almost flat out to $\deltagmtwo
\lsim 10\times 10^{-10}$, which means that the $\tau$ decay data
determination is perfectly compatible with all other
observations. Indeed, the $\test$ results in the bottom part of
table~\ref{tab:ltest1D} show that the results are not significant for
the $\tau$ decay data.  However, the predictive probability drops
fairly steeply above that value (compare
fig.~\ref{fig:1D_predictive}), thus leading to tension for the $\epem$
data, at about the $2\sigma$ level for both the flat and the CCR mSUGRA
prior. The significance is reduced to $\sim 1.6\sigma$ under the log
prior.

It is interesting how the predictive probabilities are almost
independent on the choice of priors on the model's parameters, thus
indicating a remarkable robustness in the model's predictions.  

\begin{table}
\centering
 \begin{tabular}{|l | l | l|} 
\hline
 Prior & $\ln \predlike(\deltagmtwo, \bsg |D)$ & Interpretation \\\hline\hline
 \multicolumn{3}{|l|}{$\deltagmtwo$ from $\epem$ data} \\\hline
 Flat &  $-5.99 \pm 0.13 $ & Incompatible at 95.0\% significance   \\
 Log & $ -5.87 \pm 0.13 $ & Incompatible at   94.7\% significance \\
  CCR mSUGRA &  $-6.42 \pm 0.14$ & Incompatible at  96.0\% significance \\\hline
 \multicolumn{3}{|l|}{$\deltagmtwo$ from $\tau$ decay data} \\\hline
 Flat &  $-1.59 \pm 0.07 $ &   Not significant ($1.26\sigma$)\\
 Log & $  -1.70 \pm 0.07 $ &  Not significant ($1.30\sigma$)\\
 CCR mSUGRA &  $-1.62 \pm 0.07$ & Not significant ($1.27\sigma$) \\
 \hline
\end{tabular}
\caption{Results of the $\ltest$, jointly testing $\bsg$ and
  $\deltagmtwo$ for mutual compatibility and compatibility with all
  other observations, $\dass$. We find that the $\deltagmtwo$ observation from $\epem$ data is incompatible with $\bsg$ at the $\sim 95\%$ level, almost independently of the choice of prior. On the other hand, no significant tension is detected for the $\deltagmtwo$ measurement from $\tau$ decay data. \label{tab:Ltest}  } 
\end{table}

\begin{figure}[tbh!]
\begin{center}
\begin{tabular}{c c c}
\includegraphics[width=\ttr]{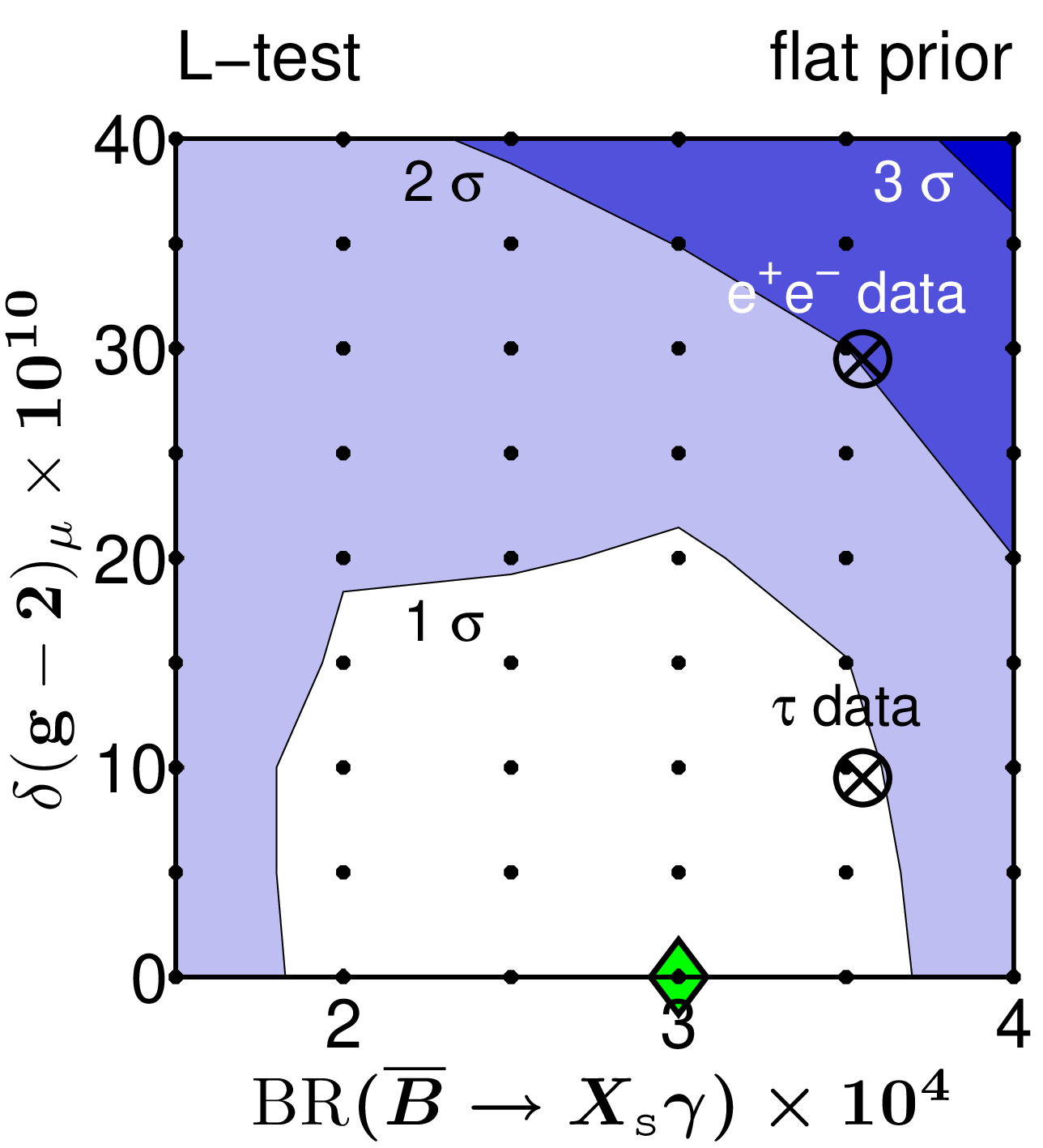}
& \includegraphics[width=\ttr]{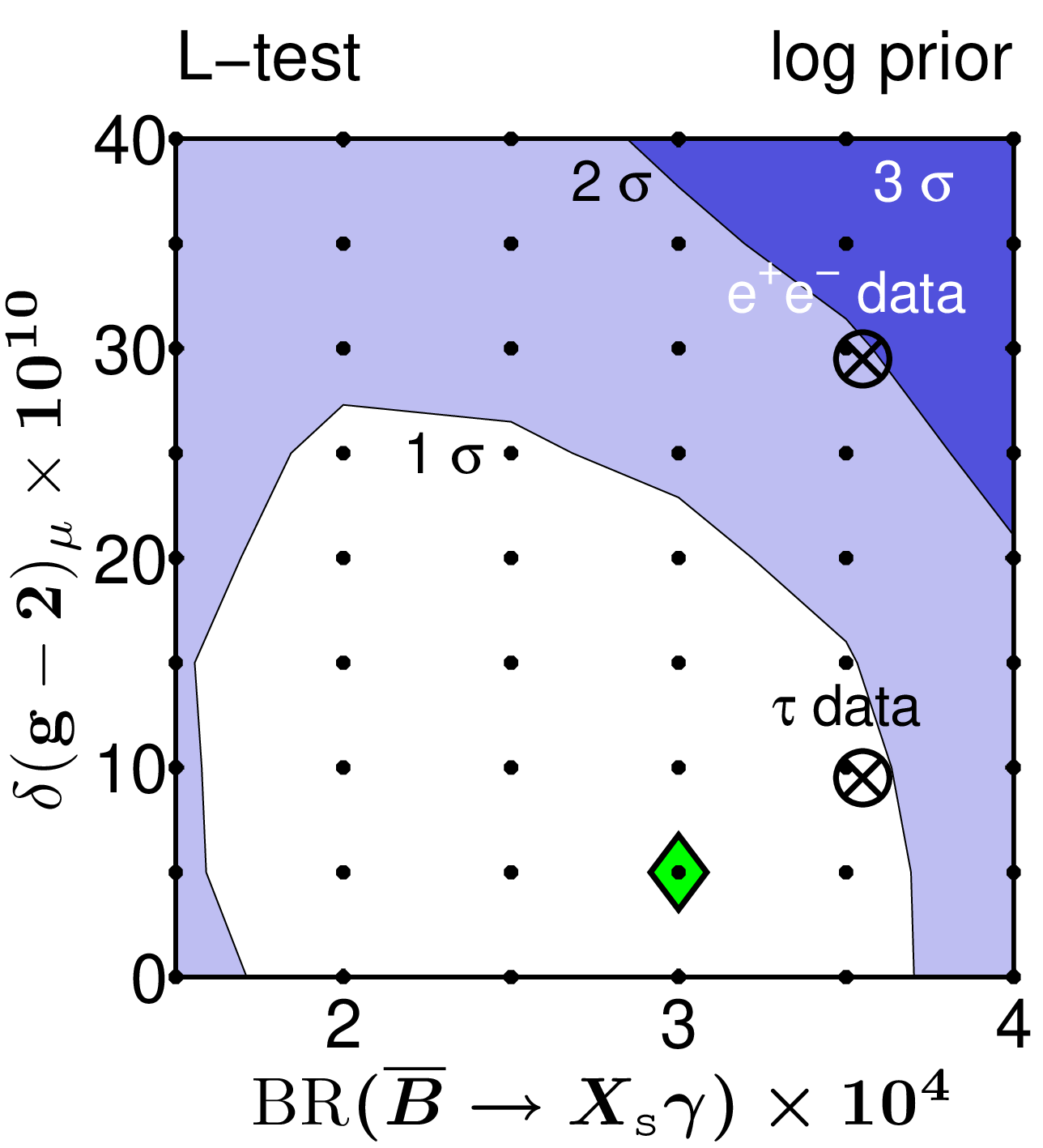}
 & \includegraphics[width=\ttr]{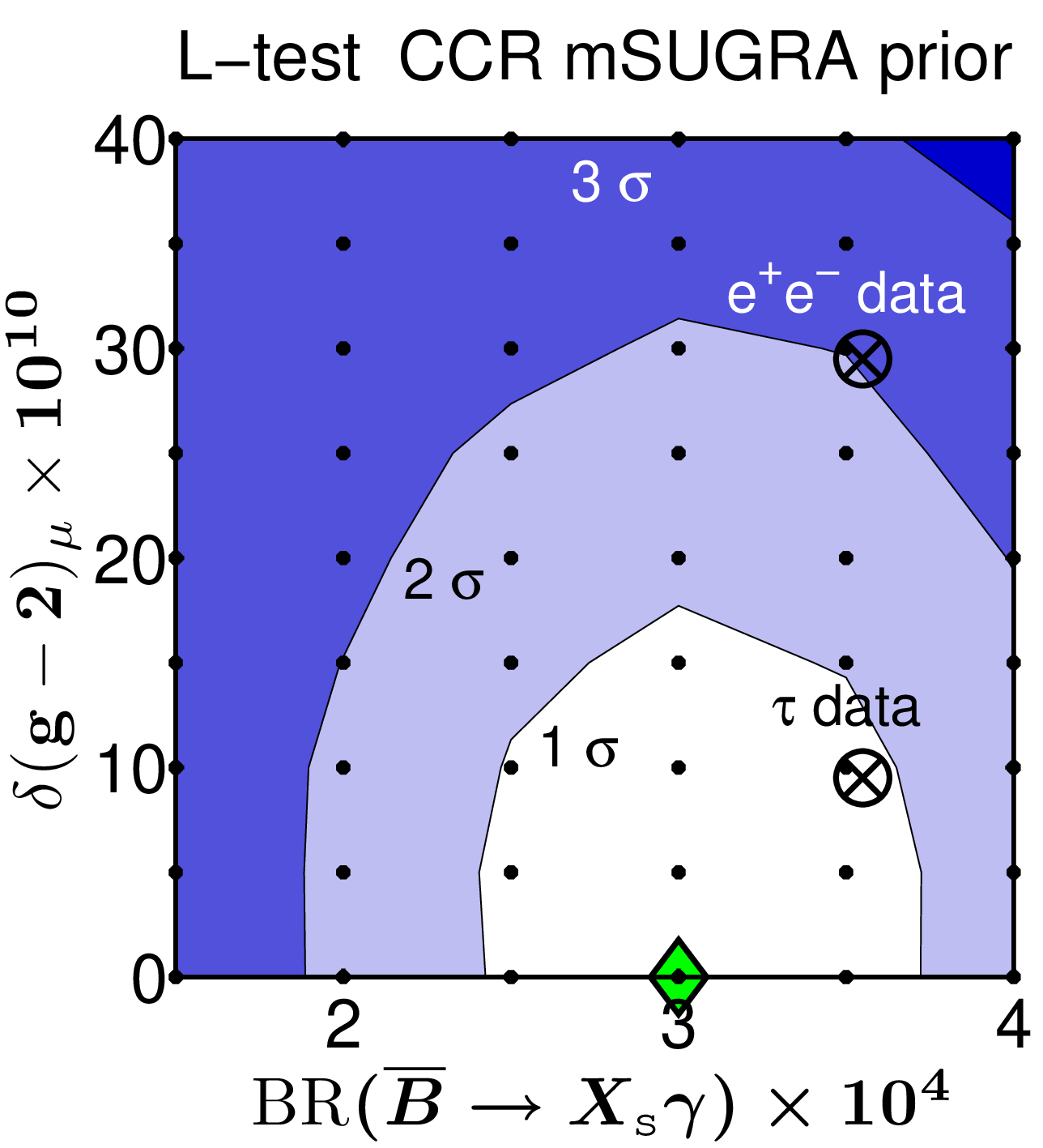}\\
\end{tabular}
\caption[test]{$\ltest$ for both $\bsg$ and
  $\deltagmtwo$, for flat priors (left panel), log priors (middle
  panel) and CCR mSUGRA priors (right panel) in the CMSSM.  The cross
  give the actual observed values (for 
  the two different $\deltagmtwo$ determinations) and the green
  diamond is the most probable value under the model. Contours delimit
  values of $\ln \predlike(\deltagmtwo, \bsg |D) = 2.3, 6.17, 11.80$,
  corresponding to joint $1,2,3\sigma$ significance regions. The
  black, small dots indicate the locations at which the predictive
  probability has been evaluated, while the contours are
  interpolated.} 
\label{fig:2D_predictive}
\end{center}
\end{figure}

We now consider the case where we test both $\bsg$ and $\deltagmtwo$
jointly, conditional on all other data. The result for the $\ltest$
with $\test = (\bsg, \deltagmtwo)$ is shown in
fig.~\ref{fig:2D_predictive} and reported in table~\ref{tab:Ltest}. We
can see that the CMSSM, given the observations $\dass$, tends to
prefer small corrections to $\deltagmtwo$, although less so in the
case of the log
prior which gives more weight to the low-mass region where SUSY
corrections tend to be larger. The joint observation of $\bsg$ and
the determination of $\deltagmtwo$ based on $\tau$ decay data lies
within the $1\sigma$ region, and hence no significant tension is
detected between these two datasets. However, the $\epem$ data based
determination of $\deltagmtwo$ shows a $\sim 2\sigma$ significance for
incompatibility (compare top part of table~\ref{tab:Ltest}), which
indicates an emerging tension between the two observations. The
r.h.s.~panel of fig.~\ref{fig:2D_predictive} gives the result for the
CCR mSUGRA prior, which as it is shown in ref.~\cite{ccr08} prefers
the focus point region and large gaugino masses, penalizing large
$\tanb$ values. This implies that, under this prior, regions of
parameter space are favoured where the decoupling of SUSY
contributions to $\deltagmtwo$ occurs and where negative contributions
of the chargino-stop loop to $\bsg$ are suppressed. (Notice how the
$3\sigma$ significance region in this case is much more
extended). Despite this, the above results hold essentially unchanged
even for this choice of prior.

\subsection{Results for the $\rtest$}

\begin{figure}[tbh!]
\begin{center}
\begin{tabular}{c c c}
\includegraphics[width=\ttr]{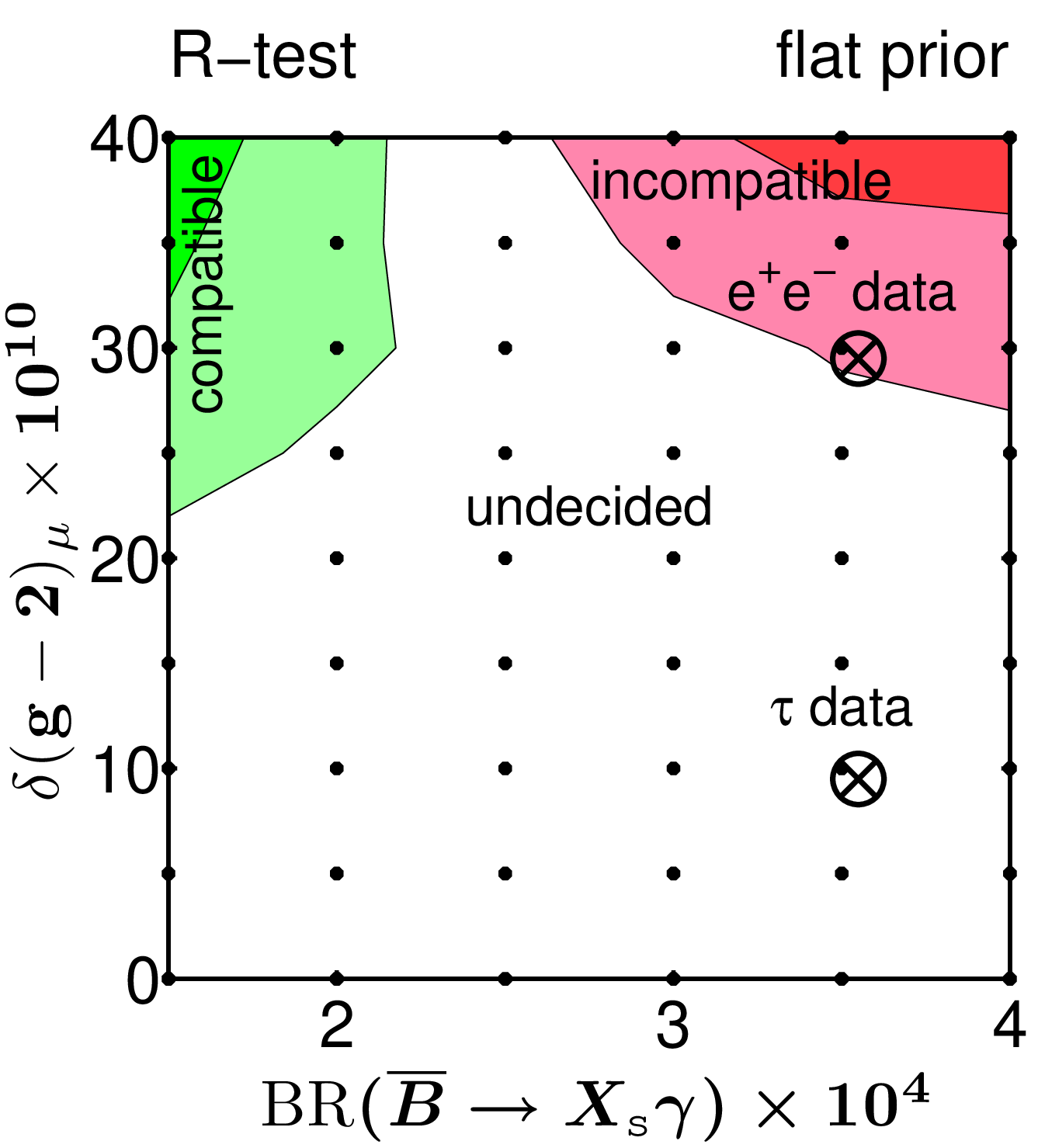}
 & \includegraphics[width=\ttr]{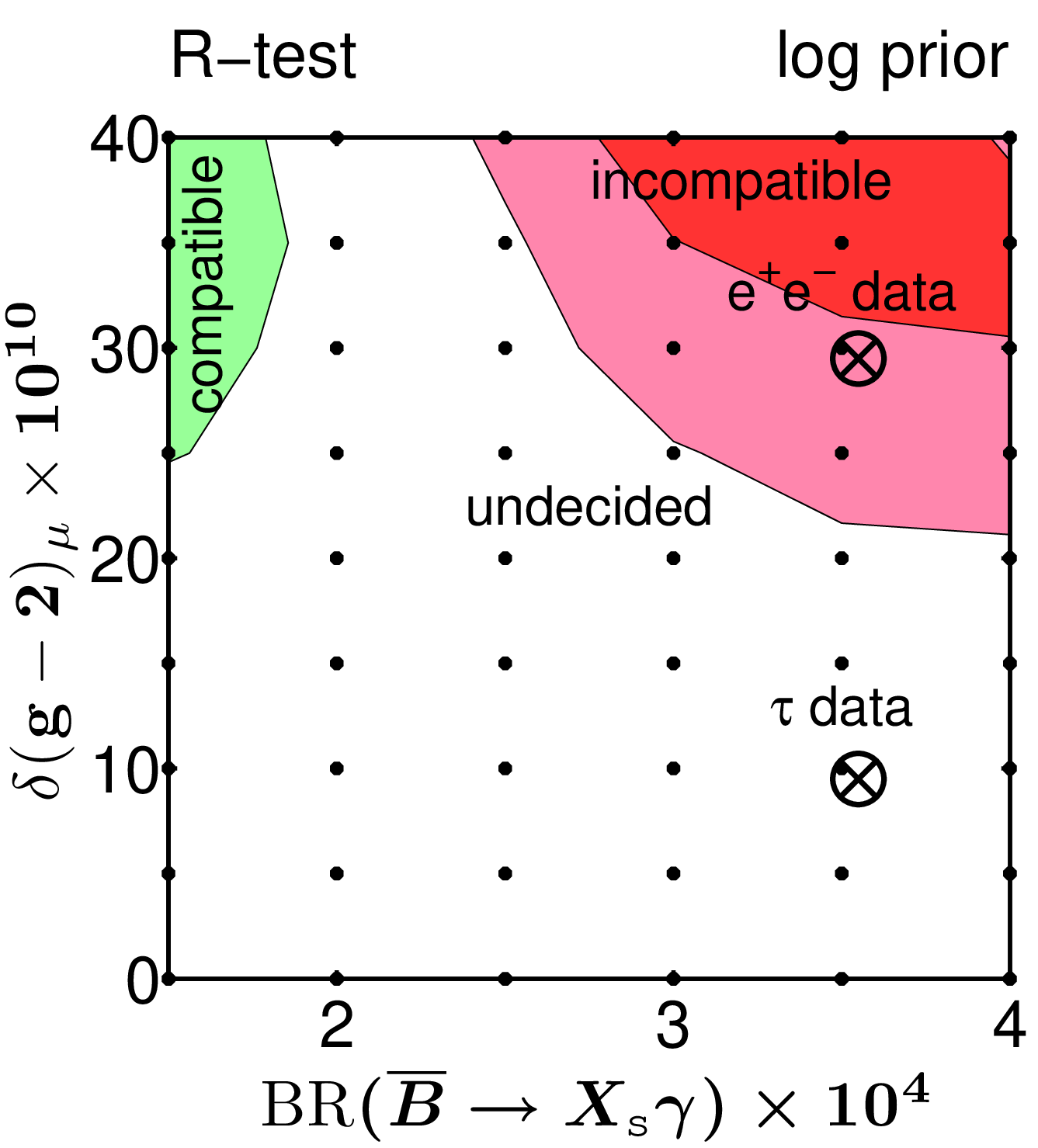}
 & \includegraphics[width=\ttr]{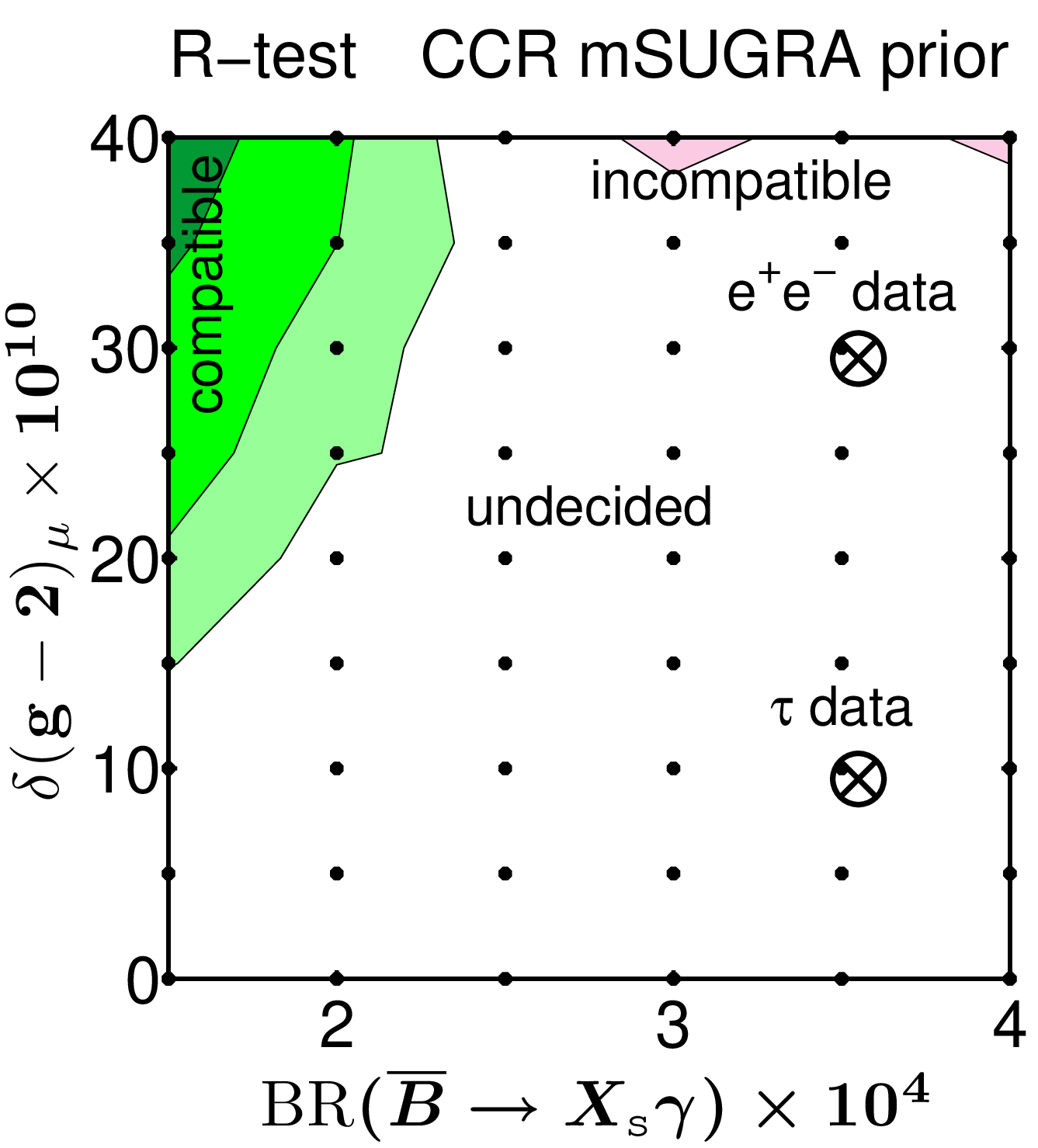}\\
\end{tabular}
\caption[test]{$\rtest$ for both $\brbsgamma$ and $\deltagmtwo$, for
  the flat prior (left panel), log prior (middle panel) and CCR mSUGRA
  prior (right panel) in the CMSSM.
  The encircled crosses give the actual observed values (for the two
  different $\deltagmtwo$ determinations). Contours delimit values of
  $|\ln \Rr (\deltagmtwo, \bsg |D)| = 1.0, 2.5$, corresponding to
  levels of weak and moderate strength of evidence for either
  hypothesis, respectively, according to the Jeffreys' scale. The
  region in the top left corner favours $\Hy_0$ (that the two
  measurements are compatible), while the top right corner favours
  $\Hy_1$ (incompatible measurements). The white region returns an
  undecided result.} 
\label{fig:2D_rtest}
\end{center}
\end{figure}

Turning now to the $\rtest$, we summarize the results in
table~\ref{tab:Rtest} and plot the outcome in
fig.~\ref{fig:2D_rtest}. The $\rtest$ for $\bsg$ and $\deltagmtwo$ returns
an inconclusive result for all choices of priors, except for the case
of log prior and $\deltagmtwo$ based on $\epem$ data alone, which instead shows
weak evidence for incompatibility. The reason for this result can 
easily be understood by considering fig.~\ref{fig:2D_rtest}, which shows
that for almost all possible observed values of $\bsg$ and
$\deltagmtwo$ is undecided. Regions of large positive SUSY contributions 
to $\deltagmtwo$ and large negative corrections to $\bsg$ would be
favoured (upper left corner of fig.~\ref{fig:2D_rtest}), while regions
of large positive corrections to $\bsg$ and large $\deltagmtwo$ values
are disfavoured (upper right corner). The relative size of those
region is somewhat prior dependent. This comes about in an analogous
fashion as for the toy model presented in the Appendix: the $\rtest$
deems observed values to be compatible if they tend to come from
``compensating'' regions of parameter space. However, by comparing
fig.~\ref{fig:2D_rtest} with
fig.~\ref{fig:2D_predictive}, it is 
apparent that in this context the $\ltest$ is the more stringent of
the two, while the $\rtest$ remains quite lenient, at least given the
current experimental error on $\deltagmtwo, \bsg$.

\begin{table}
\centering
 \begin{tabular}{|l | l | l|} 
\hline
 Prior & $\ln\Rr(\deltagmtwo, \bsg) $ &Interpretation \\\hline\hline
 \multicolumn{3}{|l|}{$\deltagmtwo$ from $\epem$ data} \\\hline
 Flat &  $-0.62 \pm 0.20 $ & Inconclusive evidence \\
 Log & $  -2.04 \pm 0.20 $ & Weak evidence for incompatibility \\
  CCR mSUGRA &  $-0.69 \pm 0.25$ & Inconclusive evidence \\\hline
 \multicolumn{3}{|l|}{$\deltagmtwo$ from $\tau$ data} \\\hline
 Flat &  $0.17 \pm 0.19 $ & Inconclusive evidence \\
 Log & $  -0.64 \pm0.17 $ & Inconclusive evidence \\
  CCR mSUGRA &  $-0.58 \pm 0.23$ & Inconclusive evidence \\
 \hline
\end{tabular}
\caption[test]{Results of the $\rtest$, giving the relative odds
  (Bayes factor) between the two hypotheses that $\deltagmtwo$ and
  $\bsg$ are mutually compatible (corresponding to $\ln\Rr > 0$) or
  that they are in tension with each other and/or the rest of the
  data, $D$ (corresponding to $\ln\Rr < 0$). The statistical
  interpretation is in accordance with Jeffreys' scale, given in
  table~\ref{tab:Jeff}. \label{tab:Rtest} }   
\end{table}

\section{Conclusions}
\label{sec:concl}
 
We have subjected the question of the mutual compatibility of  $\bsg$ and
$\deltagmtwo$ to a detailed scrutiny, employing two different statistical
tests that look for possible inconsistencies between the two
quantities and between the quantities and the model, in our case
the CMSSM. We have found no sign of tension between $\bsg$ and the
$\tau$ decay derived measurement of $\deltagmtwo$ under either
test. On the other hand, our most stringent test shows a $\sim 95\%$ indication
of tension between $\bsg$, the $\epem$--based value of $\deltagmtwo$
and the other observed data (including the WMAP 5--yr dark matter
determination). This can be interpreted in two ways: either as a sign
of undetected systematics in the $\epem$ value of $\deltagmtwo$, or
(perhaps more interestingly) as an early indication of the difficulty
of the CMSSM to simultaneously explain the observed values of
$\bsg$ and
$\deltagmtwo$. If the $\sim3\sigma$ discrepancy in the anomalous
magnetic moment of the muon is confirmed, this could be
interpreted as evidence against the viability of the CMSSM.

\medskip
{\bf Acknowledgements} \\ The authors wish to thank Louis Lyons for
many useful discussions and suggestions.  F.F. is supported by the
Cambridge Commonwealth Trust, Isaac Newton and the Pakistan Higher
Education Commission Fellowships. L.R. is partially supported by the
EC 6th Framework Programmes MRTN-CT-2004-503369 and
MRTN-CT-2006-035505. R.RDA. is supported by the project PROMETEO
(PROMETEO/2008/069) of the Generalitat Valenciana. R.T. would like to thank the Galileo Galilei Institute for
Theoretical Physics for the hospitality and the INFN for partial
support during the completion of this work.  The authors would like to
thank the European Network of Theoretical Astroparticle Physics ENTApP
ILIAS/N6 under contract number RII3-CT-2004-506222 for financial
support.  Some of the computations were carried out on the the
Cambridge High Performance Computing Cluster Darwin and the authors
would like to thank Dr.~Stuart Rankin for computational assistance.

\appendix
\section{Illustration of the consistency tests on a toy problem}
\label{appendix}

In order to illustrate the use of the Bayesian evidence to quantify
the consistency between different datasets as discussed in
section~\ref{sec:theory}, we apply the $\ltest$ (defined in
eq.~\eqref{eq:ltest}) and the $\rtest$ (defined in
eq.~\eqref{eq:rtest}) to the simple linear problem of fitting a
straight line through a set of data points, and then check for the
consistency of a new observation with the previous measurements and
with the model. 

\subsection{Toy problem}
We consider that the true underlying model for some process is a straight line described by
\begin{equation}
y( x )= m x + c,
\label{eq:line}
\end{equation}
where the free parameters in the model are the slope $m$ and the
intercept $c$, whose true value is assumed to be $1$ for both. The
data consist of observations $y_i$ at known locations $x_i$, with
Gaussian noise of known variance $\sigma$
\be
y_i-y(x_i) = \epsilon \sim \mathcal{N}(0, \sigma). 
\ee
We split the full dataset $\data$ in two parts, $\data = \{\test,
\dass\}$, and we wish to test for the consistency of the subset
$\test$ with the assumed subset $\dass$ and with the model of
eq.~\eqref{eq:line}. The likelihood function can 
then be written as
\begin{equation}
\mathcal{L}(m,c) \equiv p(\data | m,c ) = \prod_{i}^{} p(\data_i |m,c),
\label{eq:app:like1}
\end{equation}
where
\begin{equation}
p(\data_i | m,c )=\frac{1}{\sqrt{2\pi \sigma^2}}\exp[-\chi_i^2/2]
\label{eq:app:like2}
\end{equation}
and
\begin{equation}
\chi_i^2=\sum_{j}^{} \frac{(y(x_i)-y_i)^2}{\sigma^2},
\label{eq:like3}
\end{equation}
where $y(x_i)$ is the predicted value of $y$ at a given $x_i$ as a
function of $c,m$ and $y_i$ is the measured value. We impose uniform,
$\mathcal{U}(-5,5)$ priors on both $m$ and $c$.

\subsection{Consistency tests for one observables at the time}

In the first case, we wish to test for the consistency of one new
observation with a set of previously gathered data points. We take the
data set $\dass$ to consist of $9$ data points at $x = \{ 0, 1, 2, 3,
4, 5, 6, 7, 8 \}$ while the data set we wish to test, $\test$,
consists of one observation at $x_9 = 9$. For definiteness, we use
$\sigma=0.5$ for the noise. We now employ the $\ltest$ and the
$\rtest$ to check for consistency between datasets $\dass$ and $\test
= y_9 \equiv y(x_9)$.  We scan over the following $y_9$ values for $x
= 9$
\begin{equation}
\test:\qquad y_9 = 7.5, \dots, 12.5 ~~{\rm in~ intervals~ of~ 0.5}.
\end{equation}

\begin{figure}[tbh!]
\begin{center}
\includegraphics[width=\ww]{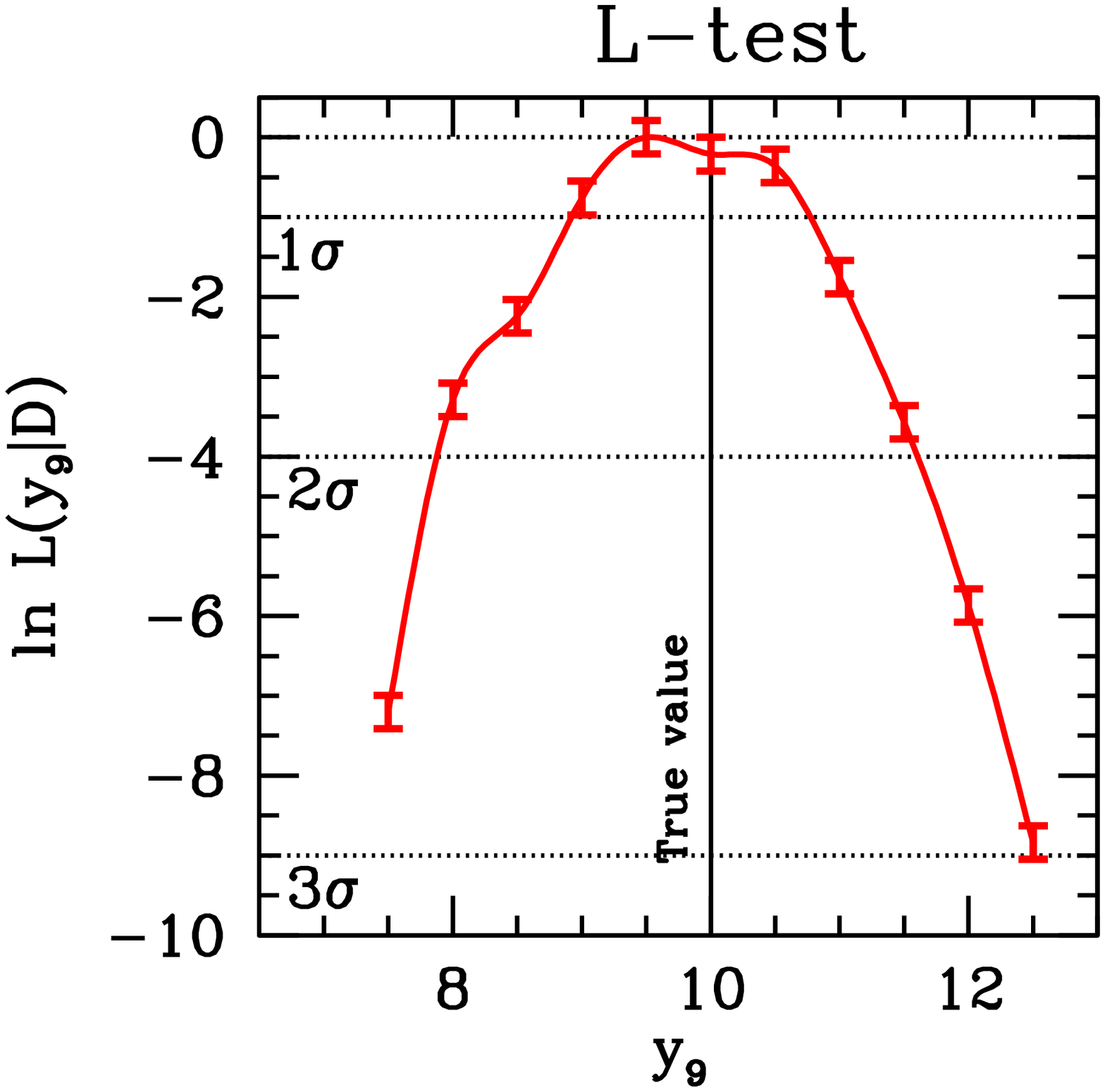} 
\includegraphics[width=\ww]{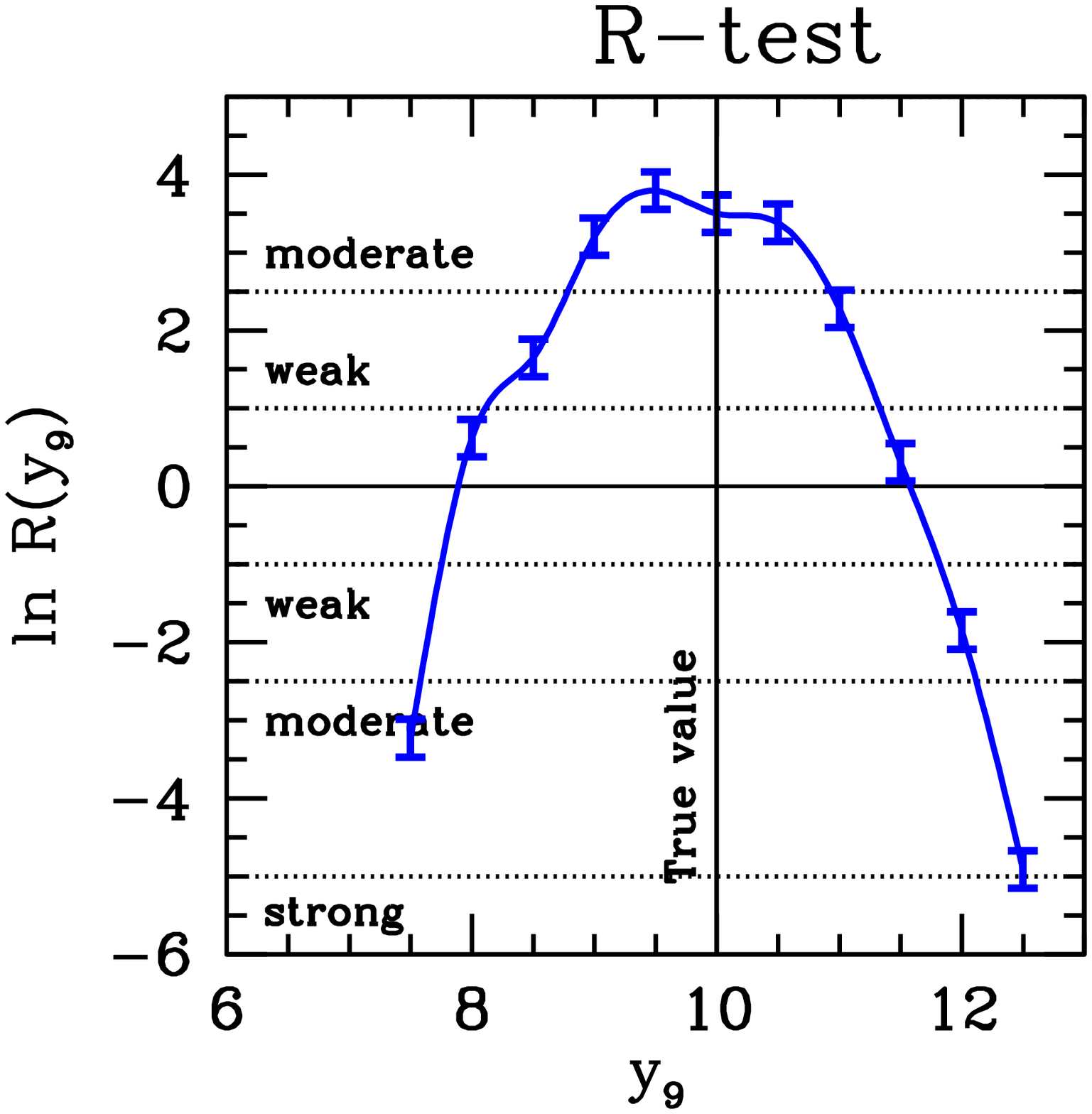}
\caption[test]{Left panel: toy model illustration of the
  1--dimensional $\ltest$ for $y_9$, with the horizontal, dashed lines
  representing levels of 1,2,3$\sigma$ significance. The vertical line
  is the value that corresponds to the true value of the model's
  parameters. Right panel: toy model illustration of the
  1--dimensional $\rtest$. The horizontal lines delineate levels of
  evidence according to the Jeffreys' scale, in favour of
  compatibility (for $\ln\Rr > 0$) or against it (for $\ln\Rr <
  0$). Both tests correctly identify the data region corresponding to
  the true model. } 
\label{fig:toys_1D}
\end{center}
\end{figure}

The outcome of these two tests is shown in fig.~\ref{fig:toys_1D},
demonstrating that both tests correctly identify the region around
$y_9 \sim 10$ as the one where the datasets are consistent (notice
that although the two curves look very similar they are not
identical). According to the $\rtest$ (right panel), the consistency
hypothesis begins to be disfavoured in the regions $y_9 \lesssim 8.0$
and $y_9 \gtrsim 11.5$, which according to the $\ltest$ corresponds to
tension between the two datasets at the $\sim 2\sigma$ level (compare
the left panel of fig.~\ref{fig:toys_1D}). As it is generically the
case when comparing hypothesis testing using likelihood ratio and
Bayesian model comparison, the significance levels of the former
appear to give stronger results than the strength of evidence from the
latter seems to justify. This is well known in the statistical
literature, and in a particular version of this phenomenon goes under
the name of ``Lindley's paradox''. For further details about
interpreting and comparing the two results,
see~\cite{Trotta:2008qt,Trotta:2005ar} and references therein.

\subsection{Consistency tests for two observables jointly}

\begin{figure}[tbh!]
\begin{center}
\includegraphics[width=\ww]{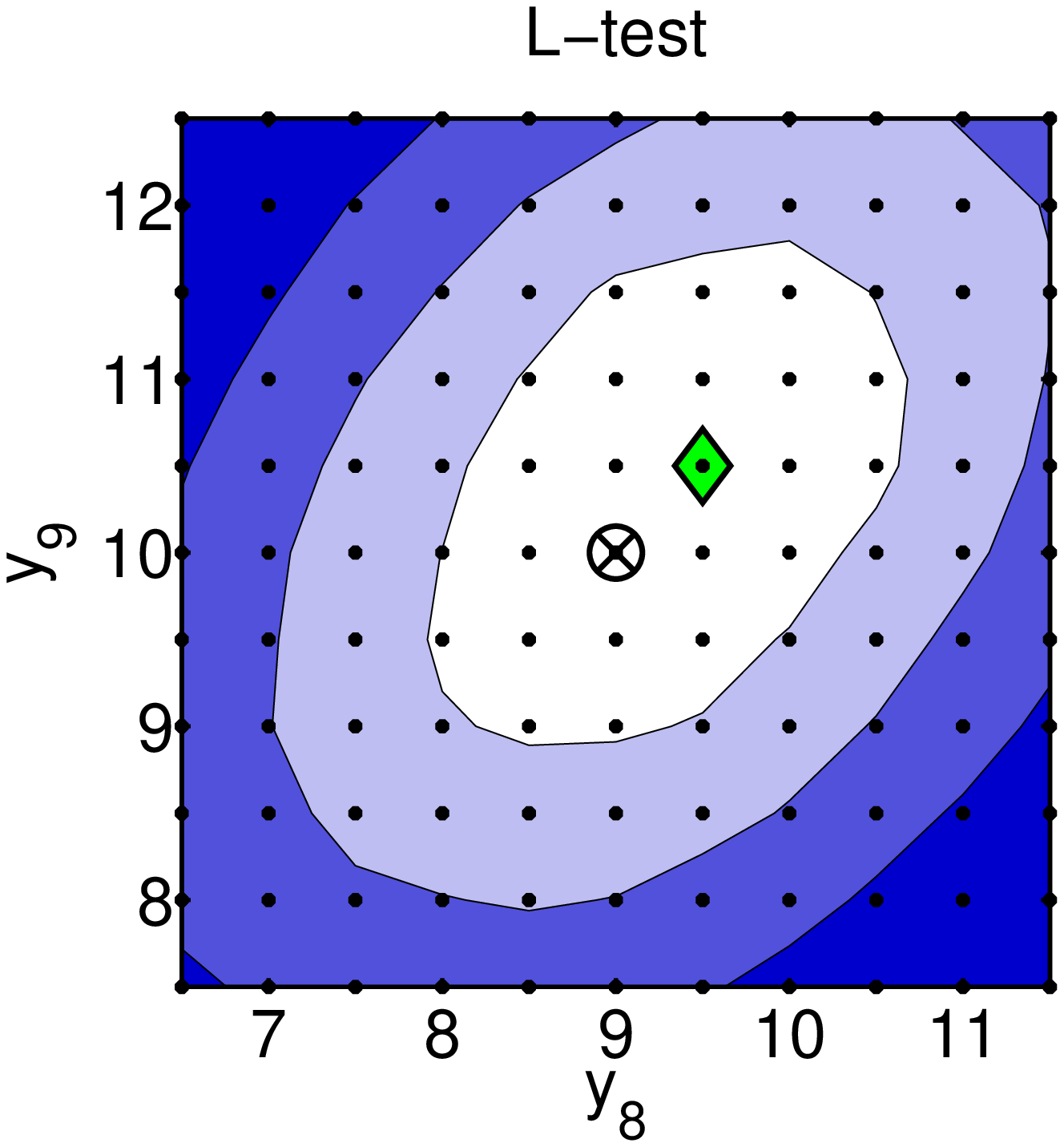} 
\includegraphics[width=\ww]{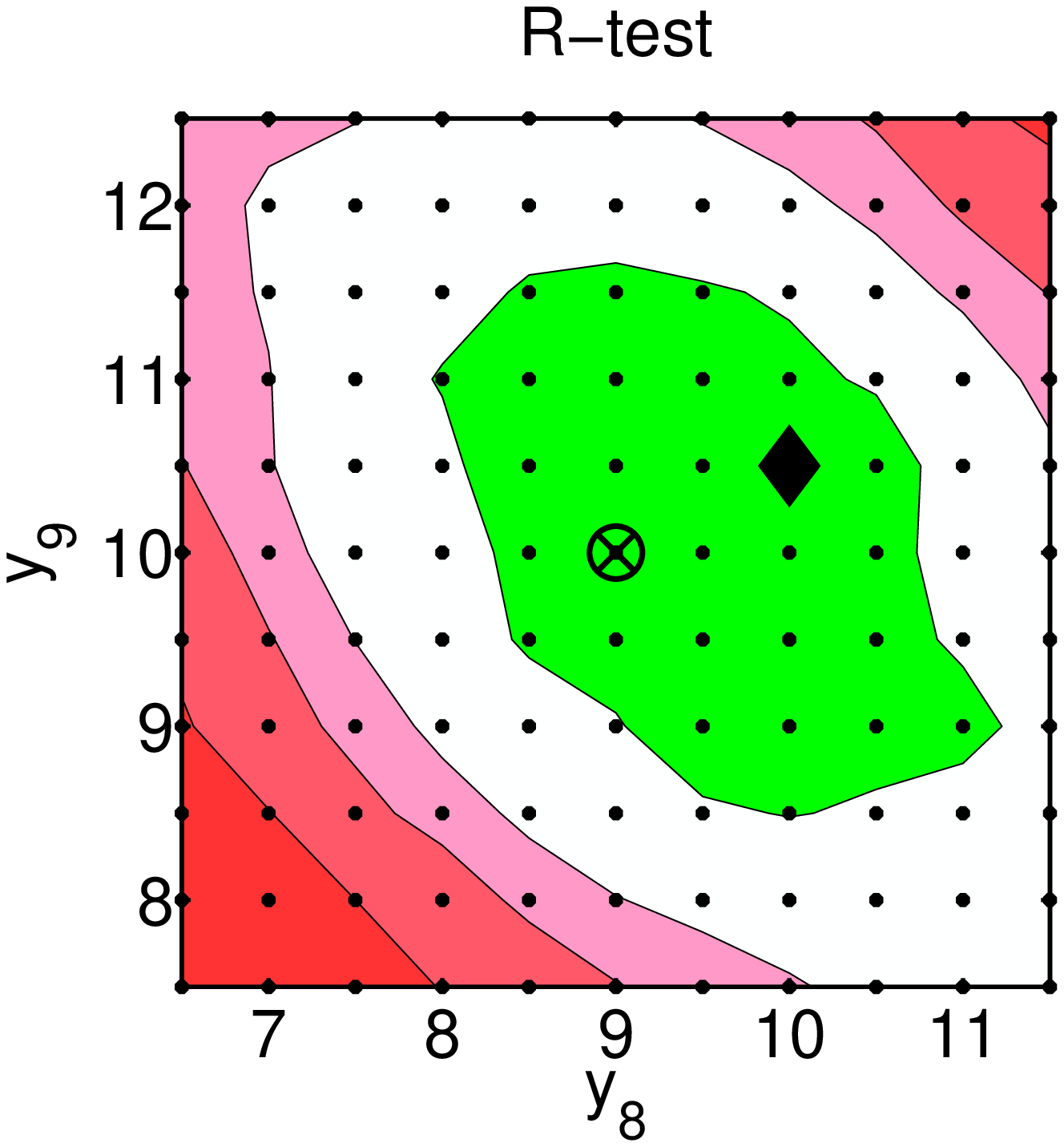}
\caption[test]{Left panel: toy model illustration of the
  2--dimensional $\ltest$, showing the true value (black, encircled
  cross), the maximum probability prediction (green diamond) and
  1,2,3$\sigma$ contours around it. Right panel: toy model
  illustration of the 2--dimensional $\rtest$. The black cross is the
  true value, while the black diamond is the point of maximum evidence
  in favour of the hypothesis of compatibility. Contours delineate
  regions of weak evidence in favour of compatibility (innermost/green
  region), inconclusive result (white), and regions of increasing
  evidence against compatibility (shades of red: weak, moderate and
  strong evidence according to the Jeffreys' scale). } 
\label{fig:toys}
\end{center}
\end{figure}

In order to perform the consistency tests for two new observations
jointly, we generate $8$ data points at $x = \{ 0, 1, 2, 3, 4, 5, 6, 7
\}$, again with Gaussian noise $\sigma=0.5$. These data points are
referred to as $\dass$. The dataset $\test$ now consists of $y(x=8)
\equiv y_8$ and $y(x=9) \equiv y_9$. We scan over the following values
for the possible outcomes of the observation for $\test$ (assuming the
same noise properties as $\dass$):
\begin{eqnarray}
y_8 = 6.5, \dots, 11.5 ~~{\rm in~ intervals~ of~ 0.5} \\\nonumber
y_9 = 7.5, \dots, 12.5 ~~{\rm in~ intervals~ of~ 0.5} .
\end{eqnarray}
The results of applying the $\ltest$ and the $\rtest$ are shown in
fig.~\ref{fig:toys}. It can be seen clearly from fig.~\ref{fig:toys}
that the both tests favour the consistency hypothesis around the
region with $y_8 \sim 9$ and $y_9 \sim 10$, which correspond to the
outcome for the true value of the parameters (marked by an encircled black
cross).

It can also be seen from fig.~\ref{fig:toys} that although there is an
overlap between different consistent and inconsistent regions favoured
by the two tests, they generally prefer different regions as they look
for inconsistency between datasets in a different manner as discussed
in section~\ref{sec:results}. For this particular model of line
fitting, the consistent region according to the $\rtest$ is the one
where the data points $y_8$ and $y_9$ lie on different sides of the
true line given by eq.~\eqref{eq:line}, i.e. the test favours either
$y_8 > 9$ and $y_9 < 10$ or $y_8 < 9$ and $y_9 > 10$. The consistent
region according to the $\ltest$ is the one where both data points
$y_8$ and $y_9$ lie on the same side of the true line, i.e. either
$y_8 > 9$ and $y_9 > 10$ or $y_8 < 9$ and $y_9 < 10$. This difference
between the two consistency tests can be understood by considering
that the $\rtest$ is trying to determine the probability that the
datasets $\test$ and $\dass$ all come from the same model and so in
order to enforce compatibility between them it favours the data points
$y_8$ and $y_9$ to lie on different sides of the true line, thus
preferring anti-correlated values. The $\ltest$, on the other hand, is
trying to fit a straight line model for the given data values and so
if $y_8$ is higher, it favours $y_9$ to be higher as well and vice
versa, therefore favouring correlated behaviour.

\appendix


\end{document}